%% file: magnonforce.tex
\documentclass[aps,prb,preprint,notitlepage]{revtex4-2}
\usepackage{amsmath}
\usepackage{graphicx}
\usepackage{amssymb}

\usepackage{gt17}
\usepackage{color} 
\usepackage{graphicx}
\usepackage{bm}
\usepackage{amsmath}
\begin{document}
\newcommand{\fv}{\bm{f}}
\newcommand{\al}{a}
\newcommand{\force}{{F}}
\newcommand{\forcecoef}{{\cal F}}
\newcommand{\magnon}{b}
\renewcommand{\dos}{{\nu_{\rm e}}}
\renewcommand{\sigmab}{\sigma_{\rm e}}
\renewcommand{\Deltasd}{\Delta}
\renewcommand{\taue}{\tau}
\title{
Calculation of magnon drag force induced by an electric current in ferromagnetic metals
}
\author{Hiroshi Funaki}
\affiliation{Kavli Institute for Theoretical Sciences, University of Chinese Academy of Sciences, Beijing, 100190, China}
\author{Gen Tatara} 
\affiliation{RIKEN Center for Emergent Matter Science (CEMS)
and RIKEN Cluster for Pioneering Research (CPR), 
2-1 Hirosawa, Wako, Saitama, 351-0198 Japan}
\date{\today}
\begin{abstract}
Magnon drag effect induced by an applied electric field in ferromagnetic metals is theoretically studied by a microscopic calculation of the force on magnons arising from magnon emission/absorption and scattering due to driven electrons.
It is shown that magnon scattering contribution dominates over the emission/absorption one in a wide temperature regime in good metals with long elastic lifetime $\taue$, as the latter has a relative suppression factor of $(\Deltasd\taue)^{-2}$ due to the electron spin flip by the magnon, where $\Deltasd$ is the $sd$ exchange interaction energy.
Spin-transfer efficiency is discussed including the magnon drag effect.
\end{abstract}

\maketitle

\section{Introduction}
Magnon or spin wave is an elementary excitation in magnetic systems, which are expected to be useful in spintronics  for information transport and processing different from the conventional electronics \cite{Chumak15}.
An obvious advantage of using magnons is the fact that they exist even in insulators, where conduction electrons are absent.
Spin-transfer effect due to magnon current was theoretically argued in the context of domain wall motion in Ref. \cite{YanMagnon11}.
Magnons are, however, not easy to control and to detect, as it does not have direct coupling to an electric field.
Thermal driving by applying a temperature gradient is therefore a common method for inducing magnon flow like in the spin Seebeck effect \cite{Uchida10Insulator}.
Unlike conduction electrons, magnon effects are generally temperature-dependent owing to a Bose distribution function for magnon excitation, but the effects are not easy to separate from other bosonic origins such as phonons with similar character.
Responses to an external magnetic field were used to identify the magnon contribution  \cite{Grannemann76}.
Magnon drag effect was experimentally identified by use of a thermopile structure to cancel nonmagnetic origins in Ref.  \cite{Costache12}.

In ferromagnetic metals, conduction electron spin is polarized due to strong $sd$ exchange interaction to the magnetization, as suggested by large magnetoresistance in magnetic multilayers and  high efficiency of the spin-transfer effect for domain walls.
Strong $sd$ exchange interaction indicates that the system is a strongly correlated fluid of magnon and electron with spin up and down, where a significant magnon drag effect by the electron flow and vice versa are expected.
In 1976, magnon-drag Peltier effect, i.e., the energy current due to magnon flow when an electric field is applied, was phenomenologically argued and magnon drift velocity was found from experimental data to be
$v_{\rm m}=\mu_{\rm m/e} v_{\rm e}$ with a constant $\mu_{\rm m/e}=2-3$, where $v_{\rm e}$ is the electron drift velocity \cite{Grannemann76}.
It is striking that the magnons are driven so efficiently by an electric field even without a direct coupling.
Magnon-drag effect was studied by use of a microscopic diagrammatic method in Ref. \cite{Miura12}.
The magnon velocity was calculated taking account of the magnon emission and absorption due to the sd exchange interaction and an applied electric field, and an expression $\mu_{\rm m/e}\propto P\eta/(\alpha_{\rm G}\Deltasd)$ was obtained for the velocity ratio, where $P$ denotes the spin polarization of conduction electron, $\eta$ is the inverse electron elastic lifetime, $\Deltasd$ is the electron spin polarization energy and $\alpha_{\rm G}$ is the (Gilbert) damping constant for magnon.
The electric current induced by the magnon current driven by a temperature gradient was calculated and argued in the context of magnonic spin-motive force in Ref. \cite{Yamaguchi19}.

\begin{figure}
 \includegraphics[width=0.4\hsize]{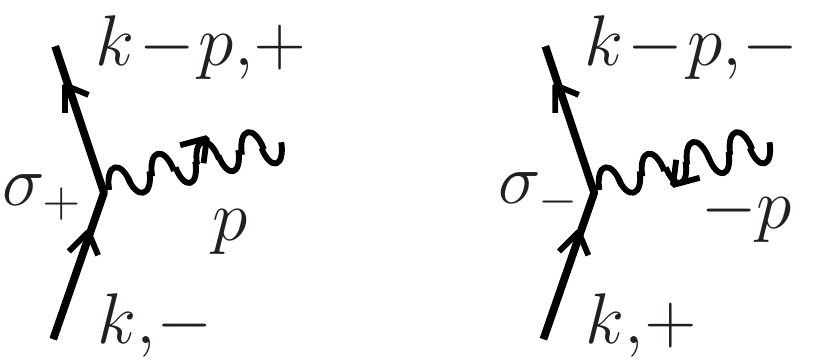}
 \includegraphics[width=0.25\hsize]{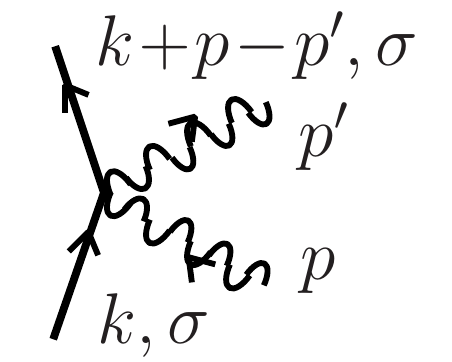}
 \caption{Diagrammatic representation of the  magnon emission, absorption and scattering.
 Solid line is the electron with spin labeled by
 $\sigma=\pm$  and wavy line is the magnon.
\label{FIGmage}}
\end{figure}

The mechanism considered in Refs. \cite{Grannemann76,Miura12}, the magnon emission and absorption, is the following process.
When the conduction electron has a drift velocity opposite to the applied electric field, a forward emission of a magnon and an absorption of a backward propagating magnon occurs
(Fig. \ref{FIGmage}). The transferred momentum from or to the magnon contributes to a drag force.
Magnon carries a spin of $-1$, and thus an absorption and emission flips the electron spin.
The absorption/emission event therefore results in a high energy excitation of the order of $2\Deltasd$ for the electron at the Fermi energy, unless the magnon wave vector matches the difference of the Fermi wave vectors, $k_{F+}-k_{F-}$ for spin $\pm$.
The electron amplitude for the force has therefore a suppression factor of $(\Deltasd\taue)^{-n}$ ($n\geq1$), where $n$ turns out to be 1 (Eq. (\ref{W1def})), $\taue$ being the electron elastic lifetime.
Being a single magnon process, the magnon amplitude for the emission/absorption turns out to depend on the temperature $T$ as $T^{\frac{5}{2}}$.
Besides the emission and absorption, electron drift causes a magnon scattering.
The scattering conserves the electron spin, and the electron stays near the Fermi energy, and thus there is no electron suppression factor for the scattering process.
The scattering is a two magnon process, whose amplitude turns out to be $\propto T^4$ and is weaker at low temperatures.
The relative strength of the emission/absorption and scattering contribution to the force is given by
$\Gamma_{{\rm ea}/{\rm sc}}\propto S^{-1}(\Deltasd\taue)^{-2}(\kb T/\tilde{J})^{-\frac{3}{2}}$
(Eq. (\ref{totalf})),  where $\tilde{J}$ is the ferromagnetic exchange energy, $S$ is the magnitude of localized spin and $\kb$ is the Boltzmann constant. The factor of $S$ is due to the fact that  emission/absorption and absorption processes corresponds to the first and the second order of the $1/S$-expansion for the magnons.
The emission/absorption process therefore dominates  for
$\kb T \lesssim S^{-\frac{2}{3}}(\Deltasd\taue)^{-\frac{4}{3}} \tilde{J}$, which correspond to a very low temperature for good metals with a long elastic lifetime.

Besides the Peltier effect, magnon drag effect contributes to other transport effects.
In fact, magnon velocity contributes to a temperature-dependent spin-transfer effect, and  magnon drag force is directly detected as a contribution to electric conductivity.
Close analysis of experimental data on various transport effects would make possible the separation of the magnon drag effect from other effects.
The aim of the paper is to provide a comprehensive theory of magnon drag effects on spin-transfer effect and resistivity by calculating the drag force microscopically using the approach of Ref. \cite{Funaki21,FunakiAH21} taking account of the magnon scattering effect.
It turns out that the magnon scattering effect for the magnon drag force and spin-transfer effect is larger than the emission/absorption effect in the wide temperature range as we argued above.

Our result for the magnon emission/absorption contribution is $\mu_{\rm m/e}\propto \eta/(\alpha_{\rm G}\Deltasd)$ (the first term of Eq. (\ref{sigmamsigmaeratio})), and is different from the result of Ref. \cite{Miura12}, which is proportional to $P$.
Comparing the analyses of Ref. \cite{Miura12} and ours, the difference appears to arose from perhaps an insufficient treatment  in Ref. \cite{Miura12} of electron (lesser) and hole (greater) contributions of the vertex function $\Lambda$ (Eq. (\ref{LambdaKar})).
Physically, the absence of $P$ in the magnon velocity appears natural as driven electrons with both spin contribute to the forward motion of magnons. In fact, up spin electron, which can only absorb a magnon, transfers positive momentum to magnons by absorbing magnons moving backward relative to the electron, while down spin electron pushes magnons forward by emitting magnons forward.
The transferred momentum and force in both cases are positive and would not change sign by a reversal of spin polarization of the electrons.

\section{Model}
The system we study is the conduction electron coupling to the localized spin (magnetization) by the $sd$ exchange interaction.
The localized spin Hamiltonian we consider is the one with an exchange interaction and an easy-axis anisotropy energy, represented by the strength $J$ and $K$, respectively:
\begin{align}
 H_{S} &= \frac{1}{2\al^3} \intr\lt[ J (\nabla\Sv)^2 - K S_z^2 \rt]
\end{align}
where $\al$ is the lattice constant.
We consider the case where the localized spin $\Sv$ is polarized along the $z$ axis.
The fluctuation around the average, the magnon, is taken into account using the Holstein-Primakov expansion as
\begin{align}
 \Sv=\lt( \begin{array}{c}
           \sqrt{\frac{S}{2}}(\magnon+\magnon^\dagger) \\
           \sqrt{\frac{S}{2}}(-i)(\magnon-\magnon^\dagger) \\
           S- \magnon^\dagger \magnon
          \end{array} \rt)  +O(\magnon,\magnon^\dagger)^3
\end{align}
The magnon energy for a wave vector $\pv$ reads
\begin{align}
 \omega_p\equiv JS p^2+KS \label{magnondispersion}
\end{align}
The current density of magnon is
\begin{align}
  \jv_{\rm m} &= -i{JS}  \magnon^\dagger \stackrel{\leftrightarrow}{\nabla} \magnon
\end{align}
and momentum density is $\pv_{\rm m}= -\frac{i}{2} \magnon^\dagger \stackrel{\leftrightarrow}{\nabla} \magnon$.

The $sd$ exchange interaction Hamiltonian is
\begin{align}
 H_{sd} &=- \frac{\Deltasd}{S} \intr \Sv\cdot(c^\dagger\sigmav c)
\end{align}
where $\Deltasd$ is the $sd$ coupling energy and $S\equiv |\Sv|$.
The interaction in terms of magnon field reads $ H_{sd} = H_{sd} ^{(1)}+H_{sd} ^{(2)}$, where
\begin{align}
 H_{sd}^{(1)} &=-\frac{\Deltasd}{\sqrt{2S}} \intr \lt[\magnon c^\dagger\sigma_- c +\magnon^\dagger c^\dagger\sigma_+ c \rt] \nnr
  H_{sd}^{(2)} &= \frac{\Deltasd}{S} \intr \magnon^\dagger \magnon c^\dagger\sigma_z c
\end{align}
represent magnon emission/absorption and scattering, respectively.

Force on magnon induced by the electron is calculated by evaluating $\dot{\pv}_{\rm m}$ taking account of $H_{sd}$.
The force due to the linear order of the $sd$ exchange interaction,
$\force^{(1)}_i \equiv  \frac{1}{2} [H_{sd}^{(1)} ,  \magnon^\dagger \stackrel{\leftrightarrow}{\nabla}_i \magnon]$ is
\begin{align}
 \force^{(1)}_i(\rv)
 &= - \frac{\Deltasd}{2\sqrt{2S}} \lt[
  (\nabla_i \magnon^\dagger)(c^\dagger \sigma_+ c) -\magnon^\dagger \nabla_i (c^\dagger \sigma_+ c)
 +(\nabla_i \magnon)(c^\dagger \sigma_- c) -\magnon \nabla_i (c^\dagger \sigma_- c) \rt]
\end{align}
Defining Fourier transform  as $c(\rv)=\sum_{\kv}e^{i\kv\cdot\rv}c_\kv$, $\magnon(\rv)=\sum_{\pv}e^{i\pv\cdot\rv}\magnon_\pv$, $\force^{(1)}_i(\rv) \equiv  \sum_\qv e^{i\qv\cdot\rv} \force^{(1)}_i(\qv)$, the momentum representation is
\begin{align}
 \force^{(1)}_i(\qv)
 &= -i\frac{\Deltasd}{2\sqrt{2S}} \sum_{\kv\kv'\pv}(k'-k+p)_i \lt[
  \magnon_{-\pv}^\dagger (c_{\kv'}^\dagger \sigma_+ c_{\kv}) +\magnon_{\pv} (c_{\kv'}^\dagger \sigma_- c_{\kv})
     \rt]_{\qv=-\kv'+\kv+\pv}
\end{align}
The uniform component of the force is
\begin{align}
 \force^{(1)}_i(0)
 &= i \frac{\Deltasd}{ \sqrt{2S}}\sum_{\kv\pv}p_i \lt[
  \magnon_{\pv}^\dagger (c_{\kv-\pv}^\dagger \sigma_+ c_{\kv}) +\magnon_{-\pv} (c_{\kv-\pv}^\dagger \sigma_- c_{\kv}) \rt]
\end{align}
showing that spin down electron emitting a magnon of momentum $\pv$ and spin up electron absorbing magnon of $-\pv$ result in a 'positive' force on magnon (Fig. \ref{FIGmage}).

The force due to the magnon scattering contribution is
\begin{align}
 \force^{(2)}_i(\rv)
 &= - \frac{\Deltasd}{S}  \nabla_i (c^\dagger \sigma_z c) \magnon^\dagger \magnon
\end{align}
indicating that the force arises from the compression of the electron spin density.
The uniform component in the Fourier representation is 
\begin{align}
\force^{(2)}_i(\qv=0)
 &= -i{\Jsd}\sum_{\kv\pv\pv'}(p'-p)_i
  \magnon_{\pv'}^\dagger \magnon_{\pv} (c_{\kv'}^\dagger \sigma_z c_{\kv})|_{\kv'=\kv+\pv-\pv'}
  \label{f2unif}
\end{align}

\section{Linear response calculation of magnon force}

We calculate the expectation value of magnon force as a linear response to an applied electric field, represented by use of a vector potential, $\Av$, where $\Ev=-\dot{\Av}$. The coupling to the electric current is represented by an interaction Hamiltonian
\begin{align}
 H_{\rm em} &= - \sum_{\kv\qv} \frac{e}{m}\lt(\Av(\qv,t)\cdot \kv\rt)  c_{\kv+\frac{\qv}{2}}^\dagger c_{\kv-\frac{\qv}{2}}
\end{align}
where $e$ is the electron charge.

\subsection{Magnon emission/absorption contribution}

\begin{figure}
 \includegraphics[width=0.8\hsize]{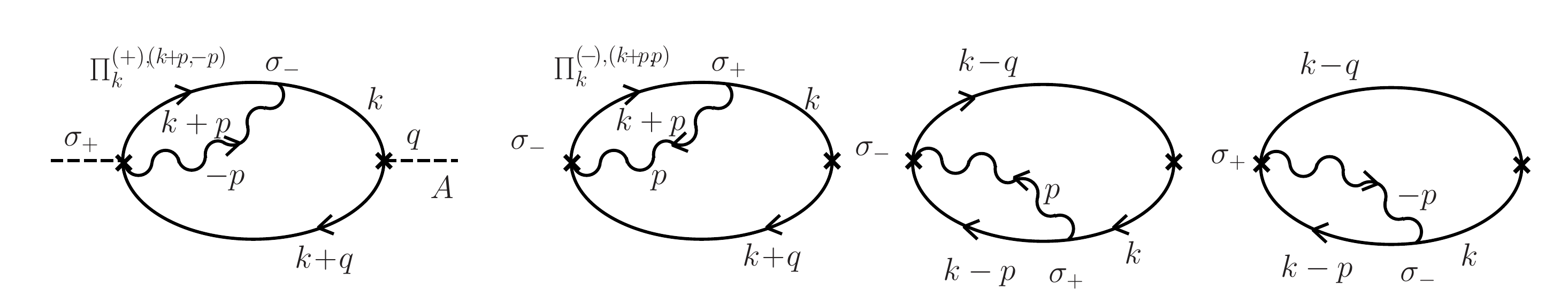}
 \caption{
 Feynman diagrams for the force on magnon due to emission and absorption.
The external wave vector, $\qv$, is zero for the uniform component we consider.
\label{FIGmagnonforce_1}}
\end{figure}

\begin{figure}
 \includegraphics[width=0.4\hsize]{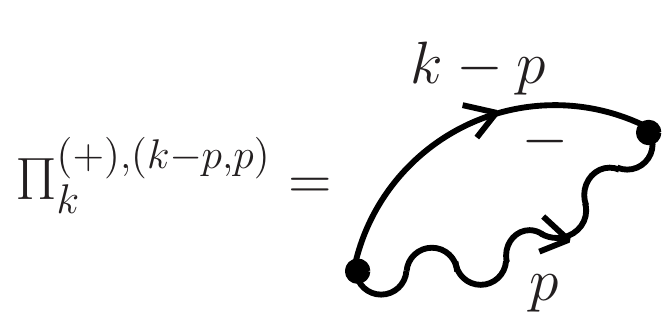}
 \includegraphics[width=0.4\hsize]{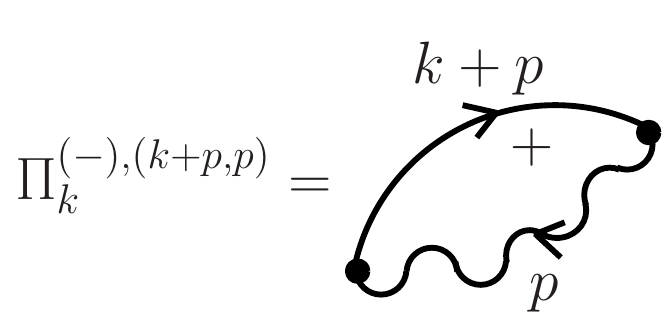}
 \caption{Diagrammatic representation of the two magnon-electron pair propagators. Electron spin is $+$ for a magnon-electron pair and $-$ for a magnon hole-electron pair.
\label{FIGpairpropagators}}
\end{figure}

In this subsection, the magnon force due to emission and absorption is calculated to the lowest (the second) order in the $sd$ exchange interaction.
We focus on the uniform component of the force.
Using  the path-ordered (non-equilibrium or Keldysh) Green's function, $g_{\kv}(t,t')$ ($\kv $ is the electron wave vector),  defined for time on a path $C=C_\rightarrow+C_\leftarrow$ (See Appendix \ref{SECAPPmepropagator}), the expectation value of the force is written in terms of the lesser component for the electron contribution as (diagrams shown in Fig. \ref{FIGmagnonforce_1})
\begin{align}
 \force^{(1)}_i(t,\qv)
  =&
  \frac{e{\Deltasd}^2}{2Sm}\sum_{\kv\kv'\pv}\sumOm A_j(\Omega)  p_i k_j\int_C dt_1\int_C dt_2 \tr \nnr
 & \biggl[
   e^{-i\Omega t_2}  \sigma_-
  \Pi_{\kv}^{(+),(\kv-\pv,\pv)}(t,t_1)  \sigma_+ g_{\kv}(t_1,t_2) g_{\kv}(t_2,t') \nnr
  & - e^{-i\Omega t_1}   \sigma_+ g_{\kv}(t,t_1)  g_{\kv}(t_1,t_2) \sigma_- \Pi_{\kv}^{(+),(\kv-\pv,\pv)}(t_2,t')\nnr
&  + e^{-i\Omega t_1}   \sigma_- g_{\kv}(t,t_1)g_{\kv}(t_1,t_2) \sigma_+ \Pi_{\kv}^{(-),(\kv+\pv,\pv)}(t_2,t') \nnr
  & -  e^{-i\Omega t_2} \sigma_+ \Pi_{k}^{(-),(\kv+\pv,\pv)}(t,t_1)
   \sigma_- g_{\kv}(t_1,t_2)g_{\kv}(t_2,t')
  \biggr]
\end{align}
Here the time $t$ is on the path $C_\rightarrow$ while $t'$ is on the path $C_\leftarrow$, and they correspond to the same real time $t$.
We used the fact that the magnon operator at the force vertex can be at either $t$ or $t'$.
The magnon-electron composite propagators are defined as (Fig. \ref{FIGpairpropagators})
\begin{align}
 \Pi_{k}^{(+),(k-p,p)}(t,t') & \equiv i g_{k-p}(t,t') d_p(t,t') \nnr
 \Pi_{k}^{(-),(k+p,p)}(t,t') & \equiv i g_{k+p}(t,t') d_p(t',t)
 \label{Pidefs}
\end{align}
corresponding to the propagation in the same ($+$) and opposite ($-$)time direction, respectively, where $d_p(t,t')$ is the magnon Green's function.
The imaginary factor $i$ is for
 $\Pi_{k}^{(\pm),(k-p,p)}$ to have the same behavior like a single Green's function, such as the lesser and greater components are pure imaginary.
%
Writing in terms of the real-time Green's functions using properties of the Green's functions and composite propagators described in Appendix \ref{SECAPPmepropagator},
the force reads
\begin{align}
 \force^{(1)}_i
 =& \frac{e\Deltasd^2}{2Sm} \sum_{\kv\kv'\pv} \sumOm \sumom A_j(\Omega)  p_i k_j  \tr \biggl[ \nnr
 & [\sigma_-
 \Pi_{k\omega}^{(+),(k-p,p),\ret} \sigma_+ g_{k\omega}^\ret g^<_{k,\omega-\Omega} +(\ret<\adv)+(<\adv\adv)]
 \nnr
 &-  [\sigma_+ g_{k,\omega+\Omega}^\ret g_{k\omega}^\ret \sigma_- \Pi_{k\omega}^{(+),(k-p,p),<}+(\ret < \adv)+(<\adv\adv)]
 \nnr
 & -  [ \sigma_+ \Pi_{k\omega}^{(-),(k+p,p),\ret}\sigma_- g_{k\omega}^\ret  g_{k,\omega-\Omega}^< +(\ret<\adv)+(<\adv\adv)]
 \nnr
 &+ [\sigma_- g_{k,\omega+\Omega}^\ret g_{k\omega}^\ret \sigma_+ \Pi_{k\omega}^{(-),(k+p,p),<} +(\ret < \adv)+(<\adv\adv)]
 \biggr]
\end{align}
In the case of static electric field ($\Omega\ra0$), the Fermi surface contribution, which is dominant, is obtained using the relation (\ref{Pistaticrelation}) as
\begin{align}
 \force^{(1)}_i
 =& \frac{e\Deltasd^2}{2Sm} \sum_{\kv\kv'\pv} \sumOm \sumom  \Omega A_j(\Omega)  f'(\omega) p_i k_j\tr \biggl[ \nnr
 &  \lt[\sigma_-
 \Pi_{k\omega}^{(+),(k-p,p),\ret} \sigma_+ g_{k\omega}^\ret g^\adv_{k,\omega}
 -  \sigma_+ g_{k,\omega}^\ret  g_{k,\omega}^\adv \sigma_- \Pi_{k\omega}^{(+),(k-p,p),\adv}
 \rt]\nnr
 &-\lt[
  \sigma_+ \Pi_{k\omega}^{(-),(k+p,p),\ret}\sigma_- g_{k\omega}^\ret  g_{k,\omega}^\adv
-
  \sigma_- g_{k,\omega}^\ret  g_{k\omega}^\adv \sigma_+ \Pi_{k\omega}^{(-),(k+p,p),\adv}
 \rt]
 \biggr]
\end{align}
where $f(\omega)\equiv [e^{\beta\omega}+1]^{-1}$, $\beta\equiv (\kv T)^{-1}$, $\kb$ and $T$ being the Boltzmann constant and temperature, respectively.
At low temperature, $f'(\omega)\simeq -\delta(\omega)$, and we obtain uniform component as
$ \force^{(1)}_i=\forcecoef^{(1)}_{ij}eE_j$
($E=\int\frac{d\Omega}{2\pi}e^{-i\Omega t}i\Omega A_{\Omega}$), where the coefficient is
($g_{\kv}\equiv g_{\kv,\omega=0}$)
\begin{align}
 \forcecoef^{(1)}_{ij}
 =&
 -\frac{\Deltasd^2}{4\pi Sm} \sum_{\kv\kv'\pv}  p_i k_j    \Im \tr \biggl[
  \sigma_-  |g_{k}^\adv|^2 \sigma_+ \Pi_{k\omega=0}^{(-),(k+p,p),\adv}
 -
  \sigma_+ |g_{k}^\adv|^2 \sigma_- \Pi_{k\omega=0}^{(+),(k-p,p),\adv}
  \biggr]\nnr
  =&
  -\frac{\Deltasd^2}{2\pi Sm} \sum_{\kv\kv'\pv}  p_i k_j  \sum_{\pm}(\pm)
    |g_{k\pm}^\adv|^2 \Im [\Pi_{k,\mp}^{(\mp),(k\pm p, p),\adv}]
    \label{fijeq1}
\end{align}
where $\Pi_{k,\mp}$ denotes $\Pi_{k,\omega=0}$ with electron spin $\mp$.
Writing the imaginary part of the pair propagator by use of magnon propagator (Eq. (\ref{ImPi2result})), we obtain
\begin{align}
 \forcecoef^{(1)}_{ij}
  =&
  \frac{2\Deltasd^2}{\pi S m} \sum_{\kv\kv'\pv}  p_i k_j \sum_\nu  n_\nu(1-f_\nu) \sum_{\pm}\Im[d^\adv_{p,\mp \nu}]
    |g_{k\pm}^\adv|^2 \Im [g_{k\pm p,-\nu,\mp}^\adv]
\end{align}
where $\nu$ is the magnon frequency, $\sum_\nu=\int\frac{d\nu}{2\pi}$ and $n_\nu\equiv [e^{\beta\nu}-1]^{-1}$ is the Bose distribution function.
The magnon Green's function at weak damping is
\begin{align}
 \Im[d^\adv_{p,\mp \nu}]=& \Im \frac{1}{\mp\nu-\omega_p-i\eta_{\rm m}}
 \simeq \pi \delta(\mp\nu-\omega_p)
\end{align}
where $\eta_{\rm m}$ represents damping of magnon, which is
$\eta_{\rm m}=\alpha_{\rm G}\omega_p$ in terms of the Gilbert damping constant $\alpha_{\rm G}$.
To proceed, we consider the case where magnon dynamics is slow compared to the electron one, namely, frequency $\nu$ in electron Green's functions is treated as zero.
The approximation assumes therefore that $\omega_{\rm m} \taue \ll 1$, where $\omega_{\rm m}$ is typical magnon energy.
In this case, we obtain the sum of $\nu$ as
$\sum_\nu  n_\nu(1-f_\nu) \Im[d^\adv_{p,\mp \nu}] =\mp \frac{1}{2}n_{\omega_p}(1-f_{\omega_p}) $, where we used
$n_{-\nu}+f_{-\nu}= - (n_{\nu}+f_{\nu})=-2n_\nu(1-f_{\nu})$.
We thus have
\begin{align}
\forcecoef^{(1)}_{ij}
  =&
  -\frac{\Deltasd^2}{\pi Sm} \sum_{\kv\kv'\pv}  p_i k_j
  n_{\omega_p}(1-f_{\omega_p})\sum_{\pm}(\pm)
    |g_{k\pm}^\adv|^2 \Im [g_{k\pm p,0,\mp}^\adv]
\end{align}
The wave vector for magnon is also assumed to be small, $p\ll \kf$.
The expression in this case becomes, using
$g_{k\pm p,-\nu,\mp}^\adv\simeq g_{k \pm p,\mp}^\adv
\simeq g_{k\mp}^\adv\pm\frac{\kv\cdot\pv}{m}(g_{k\mp}^\adv)^2$,
\begin{align}
 \forcecoef^{(1)}_{ij}
  =&
  -\frac{\Deltasd^2}{\pi Sm^2}  \sum_{\kv\pv}  p_i k_j(\kv\cdot\pv) n_{\omega_p}(1-f_{\omega_p})\sum_{\pm}
    |g_{k\pm}^\adv|^2 \Im (g_{k\mp}^\adv)^2
    \nnr
  =& \delta_{ij} \overline{\forcecoef}^{(1)} \nnr
 \overline{\forcecoef}^{(1)}
 =&
 - \frac{\Deltasd^2}{3\pi Sm^2} \sum_{\pv}  p^2 n_{\omega_p}(1-f_{\omega_p})  \sum_{\kv}\sum_{\pm}k^2
    |g_{k\pm}^\adv|^2 \Im (g_{k\mp}^\adv)^2
\end{align}
The electron part is estimated by use of contour integration with respect to electron energy as (derivation in Appendix \ref{SECelectronsum})
\begin{align}
  \Im \sum_{\kv\sigma}k^2 |g_{k\sigma}^\adv|^2 (g_{k,-\sigma}^\adv)^2
 &=
  -\frac{9\pi}{8} \frac{m a^3 \overline{\nel}}{\Deltasd^2 \ef}
 \label{electronsumf1}
\end{align}
where $\overline{\nel}=\sum_\sigma {\nel}_\sigma$ and ${\nel}_\sigma\equiv \frac{1}{3m\al^3}\sum_{\sigma}\sigma{\dos_{\sigma} k_{\sigma}^2}$ are the total and spin-resolved electron density, respectively,
$\dos_\sigma$ and $k_{\sigma}$ being the electron density of states at the Fermi energy and the Fermi wave length for spin $\sigma$, respectively.
The force coefficient is
\begin{align}
 \overline{\forcecoef}^{(1)}
 = &
 \frac{3}{8S m a^2} \frac{a^3 \overline{\nel}}{\ef}  W^{(1)}(T)
 \label{f1result}\\
W^{(1)}(T) \equiv &  a^2
\sum_{\pv}  p^2  n_{\omega_p}(1-f_{\omega_p})
\label{W1def}
\end{align}

\subsection{Magnon scattering contribution}

\begin{figure}
 \includegraphics[width=0.4\hsize]{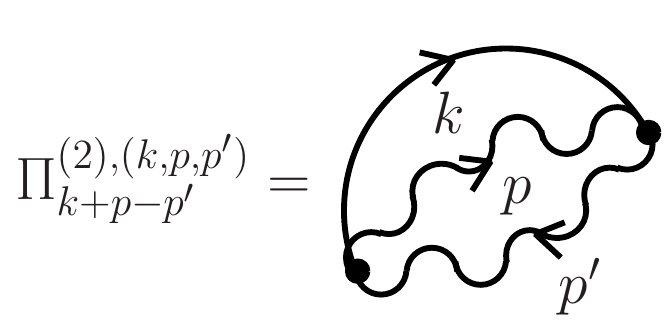}
 \caption{
 Feynman diagrams for the two magnon and electron composite propagator.
\label{FIGpairpropagator2}}
\end{figure}

The force due to the magnon scattering, $\force^{(2)}_i$, is similarly calculated.
The uniform component of the linear response at the second order in the $sd$ exchange interaction  reads
\begin{align}
 \force^{(2{\rm a})}_i(t)
 =& -\frac{e\Deltasd^2}{S^2m}   \sumOm \Av_{j}(q,\Omega) \sum_{\kv\kv'\pv}(p-p')_i k_j \int_C dt_1 \int_C dt_2 \tr
 \nnr
 &\biggl[ e^{i\Omega t_2} \sigma_z   \Pi^{(2),k',p,p'}_{k}(t,t_1) \sigma_z
   g_{\kv}(t_1,t_2) g_{\kv}(t_2,t')
  + e^{i\Omega t_2}  \sigma_z
   g_{\kv}(t,t_1)g_{\kv}(t_1,t_2) \sigma_z \Pi^{(2),k',p',p}_{\kv}(t_2,t')     \biggr]_{k'=k+p-p'}
   \label{f2expression1}
\end{align}
where a composite propagator of electron and two magnons is defined as (Fig. \ref{FIGpairpropagator2})
\begin{align}
  \Pi^{(2),k,p,p'}_{k+p-p'}(t,t') 
  &=  g_{\kv}(t,t')d_{\pv}(t,t') d_{\pv'}(t',t) 
  \label{twomagelecpropagator}
\end{align}
The leading Fermi surface term in the limit of $\Omega\ra0$ is
\begin{align}
 \force^{(2{\rm a})}_i
 =& -\frac{e\Deltasd^2}{S^2m} \sumOm \sumom \Omega \Av_{j}(\Omega) \sum_{\kv\kv'\pv}(p-p')_i k_j  f'(\omega)
 \tr  \nnr
 &\biggl[ \sigma_z   \Pi^{(2),k',p,p',\ret}_{k}(\omega) \sigma_z
   g_{\kv}^\ret(\omega) g_{\kv}^\adv(\omega)
  +   \sigma_z
   g_{\kv}^\ret(\omega)g_{\kv}^\adv(\omega) \sigma_z \Pi^{(2),k',p',p,\adv}_{\kv}(\omega)     \biggr]_{k'=k+p-p'}
\end{align}
As the Green's functions are diagonal in spin, the two Pauli matrix $\sigma_z$ becomes irrelevant.
Namely, the scattering force is a response of electron charge sector.
The effect is similar to the $z$-component of the electron magnetic susceptibility in the presence of a magnetization along $z$ direction. 
We obtain finally (see Eq. (\ref{ImPi2a}))
\begin{align}
 \force^{(2{\rm a})}_i &=\forcecoef^{(2{\rm a})}_{ij}eE_j \nnr
 \forcecoef^{(2{\rm a})}_{ij}
  &=   \frac{\Deltasd^2}{\pi S^2m}  \sum_{\pv\pv'}\sum_{\kv\sigma}(p-p')_i k_j
 \tr\biggl[ \Im(\Pi^{(2),k+p-p',p,p',\adv}_{k,\sigma,\omega=0})    |g_{\kv\sigma}^\adv|^2   \biggr]
\nnr
  &=   -2\frac{\Deltasd^2}{\pi S^2m}  \sum_{\pv\pv'}\sum_{\kv\sigma}(p-p')_i k_j \Im[d_{p\nu}^\adv]\Im[d_{p'\nu'}^\adv]
 n_\nu(1+n_{\nu'})f_{\nu'-\nu}  \Im[g_{k+p-p'}^\adv ]   |g_{\kv\sigma}^\adv|^2
\label{F2aresultPi}
\end{align}

\begin{figure}
 \includegraphics[width=0.45\hsize]{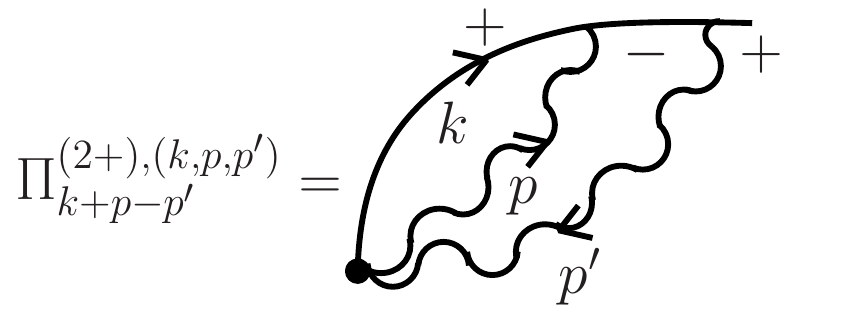}
 \includegraphics[width=0.45\hsize]{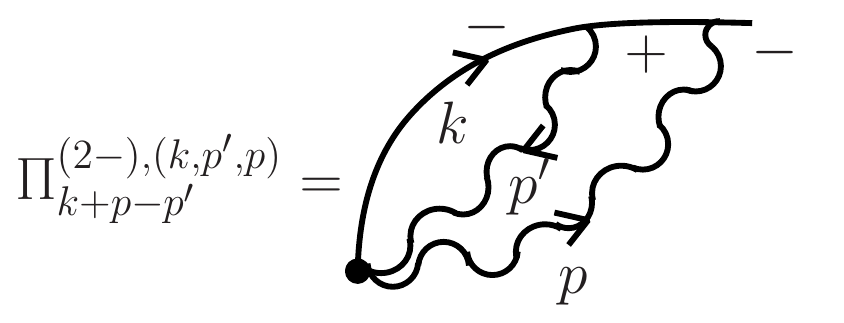}
 \caption{
 Feynman diagrams for the two magnon and electron composite propagator with magnon absorption and emission at different times. Electron spin (denoted by $\pm$) is fixed for emission and absorption processes.
\label{FIGmagnonpairpropagator2p}}
\end{figure}

The scattering force has other contributions of higher-order in the $sd$ exchange interaction, with the emission and absorption vertices at different time as in Fig. \ref{FIGmagnonpairpropagator2p}.
This contribution is characterized by the  composite propagator
\begin{align}
  \Pi_{k+p-p'}^{(2+),(k,p,p')}(t,t')
     &=
     -i\frac{\Deltasd}{2} \int_Cdt_1
     \Pi_{k+p}^{(+),(k,p)}(t,t_1) g_{k+p,-}(t_1,t') d_{p'}(t',t)
     \nnr
  \Pi_{k+p-p'}^{(2-),(k,p,p')}(t,t')
     &= -i \frac{\Deltasd}{2} \int_Cdt_1
     \Pi_{k-p'}^{(-),(k,p')}(t,t_1) g_{k-p',+}(t_1,t') d_{p}(t,t')
     \label{Pi2pm}
\end{align}
where factors are included to have the same normalization as $\Pi^{(2)}$.
The electron spin ($\pm$) is fixed for the two propagators, as the magnon absorption and emission causes lowering and highering of electron spin, respectively.
Considering a strong $sd$ splitting case, electron spin flip costs energy of  $2\Deltasd$, and thus contributions with more electron Green's functions with flipped spin are suppressed and neglected.
The electron propagators with flipped spin inside the composite propagator of Eq. (\ref{Pi2pm}) is approximated as $g_{k+p,\pm}(t,t')\sim \mp \delta(t-t')(2\Deltasd)^{-1}$, resulting in
\begin{align}
  \Pi_{k+p-p'}^{(2+),(k,p,p')}(t,t')
     & \simeq
     \frac{-i}{4}
     \Pi_{k+p,+}^{(+),(k,p)}(t,t') d_{p'}(t',t)
     =\frac{1}{4} \Pi^{(2),k,p,p'}_{k+p-p',+}(t,t')
     \nnr
  \Pi_{k+p-p'}^{(2-),(k,p,p')}(t,t')
     & \simeq \frac{i}{4}
     \Pi_{k-p',-}^{(-),(k,p')}(t,t') d_{p}(t,t')
     =-\frac{1}{4}\Pi^{(2),k,p,p'}_{k+p-p',-}(t,t')
     \label{Pi2pm2}
\end{align}
The contribution to the force coefficient from the propagators of Fig. \ref{FIGmagnonpairpropagator2p} is therefore
\begin{align}
 \forcecoef^{(2{\rm b})}_{ij}
  &=  i \frac{\Deltasd^2}{\pi S^2m}  \sum_{\pv\pv'}\sum_{\kv\sigma}(p-p')_i k_j \frac{\sigma}{4}
 \tr\biggl[ \Im(\Pi^{(2),k+p-p',p,p',\adv}_{k,\sigma,\omega=0})    |g_{\kv\sigma}^\adv|^2   \biggr]
 \label{F2bresultPi}
\end{align}

Using Eq. (\ref{ImPi2a}) and assuming $p,p' \ll k$  and low frequency magnons, we obtain the sum of the two contributions as
\begin{align}
 \forcecoef^{(2)}_{ij} &\equiv \forcecoef^{(2{\rm a})}_{ij}+\forcecoef^{(2{\rm b})}_{ij}
=\delta_{ij}\overline{\forcecoef}^{(2)} \nnr
 \overline{\forcecoef}^{(2)}
 &= - \frac{\Deltasd^2}{3S^2m^2} \sum_{\pv\pv'}(\pv-\pv')^2 n_{\omega_p}(1+n_{\omega_{p'}}) f_{\omega_{p'}-\omega_p} \sum_{\kv\sigma} k^2 \lt(1+\frac{\sigma}{4}\rt) |g_{\kv\sigma}^\adv|^2 \Im[(g_{\kv\sigma}^\adv )^2  ]
 \label{F2magresult1}
\end{align}
The summation over $\kv$ and $\sigma$ is
(using 
$k_i (g_{\kv\sigma}^\adv)^3=\frac{m}{2}\partial_{k_i} (g_{\kv\sigma}^\adv)^2$ and integral by parts)  
\begin{align}
\sum_{\kv\sigma} k^2  |g_{\kv\sigma}^\adv|^2 \Im[(g_{\kv\sigma}^\adv )^2  ]
&=-\pi m \tau^2 \sum_{\sigma}\dos_\sigma
\end{align}

We therefore obtain
($ \overline{\dos} \equiv \sum_{\sigma} \dos_{\sigma} $ and $P_\nu\equiv \frac{\sum_{\sigma} \sigma\dos_{\sigma}}{\sum_{\sigma} \dos_{\sigma}}$)
\begin{align}
 \overline{\forcecoef}^{(2)}
 = &  \frac{\pi}{3S^2ma^2} {(\Deltasd \taue)}^2  \overline{\dos}\lt(1+\frac{P_\nu}{4}\rt) W^{(2)}(T) \label{F2result} \\
W^{(2)}(T) \equiv &  a^2 \sum_{\pv\pv'}(\pv-\pv')^2 n_{\omega_p}(1+n_{\omega_{p'}}) f_{\omega_{p'}-\omega_p}
\label{f2result}
\end{align}

\subsection{Force on electron}
The recoil force on the electron with spin $\sigma$
arising from the magnon emission/absorption due to the $sd$ exchange interaction is (in the field operator form)
\begin{align}
 \force^{(1)}_{{\rm e}+,i}(\qv=0)
 &= i\frac{\Deltasd}{ \sqrt{2S}}\sum_{\kv\kv'\pv}p_i
  \magnon_{\pv} (c_{\kv'}^\dagger \sigma_- c_{\kv})_{\kv'=\kv+\pv} \nnr
 \force^{(1)}_{{\rm e}-,i}(\qv=0)
 &= i\frac{\Deltasd}{ \sqrt{2S}} \sum_{\kv\kv'\pv}p_i
  \magnon_{-\pv}^\dagger (c_{\kv'}^\dagger \sigma_+ c_{\kv})_{\kv'=\kv+\pv}
\end{align}
whose sum is opposite to the force on magnons; $ \sum_\sigma \force^{(1)}_{{\rm e}\sigma,i} = - \force^{(1)}_{{\rm m},i}$.
It turns out that the force is spin-independent, i.e.,
\begin{align}
 \force^{(1)}_{{\rm e}+{i}}
 =\force^{(1)}_{{\rm e}-{i}}
   =& -\frac{1}{2} \force^{(1)}_{i}
\end{align}
This result indicates that magnon emission and absorption induced by an electric field does not acts as spin motive force but drives only charge sector.

The force on electron spin arising from magnon scattering is calculated using
\begin{align}
 \force^{(2)}_{{\rm e}\sigma,i}
   &= \frac{\Deltasd}{S}[\nabla(\magnon^\dagger \magnon)] \sigma c^\dagger_\sigma c_\sigma
\end{align}
and its coefficient ($ \force^{(2)}_{{\rm e}\sigma}\equiv eE  \forcecoef^{(2)}_{{\rm e}\sigma}$)
is (from Eq. (\ref{F2magresult1})),
\begin{align}
 \forcecoef^{(2)}_{{\rm e}\sigma}
 &=  \frac{ {\Deltasd}^2 }{6S^2m^2}\sum_{\pv\pv'}(\pv-\pv')^2 n_{\omega_p}(1+n_{\omega_{p'}}) f_{\omega_{p'}-\omega_p} \lt(1+\frac{\sigma}{4}\rt)\sum_{\kv} k^2  |g_{\kv\sigma}^\adv|^2 \Im[(g_{\kv\sigma}^\adv )^2  ]
 \label{F2eleresult1}
\end{align}
This force is generally spin-dependent.

Those forces on the electron are different from the driven-magnon contribution to the motive force due to smooth magnetization structures  discussed in Refs. \cite{Lucassen11a,Yamaguchi19}.
In fact, the motive force in the adiabatic (slowly varying) limit are proportional to the magnon energy current linear in the magnon momentum, while the force argued in the present analysis are the second order of $p$ and $p'$, corresponding to nonadiabatic contributions.

\section{Total force on magnon}

\begin{figure}
 \includegraphics[width=0.8\hsize]{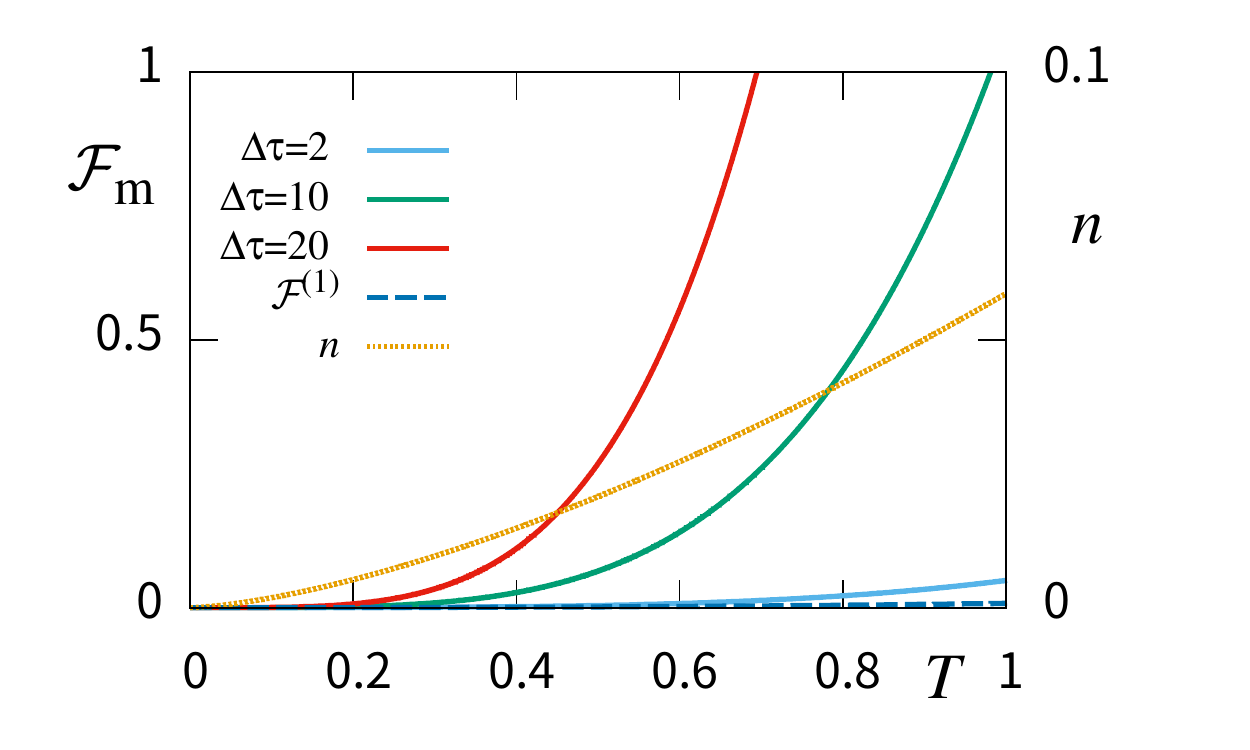}
 \caption{The total force coefficient $\forcecoef_{\rm m}$  as function of $\tilde{T}\equiv \kb T/\tilde{J}$ (denoted by $T$ in the $x$ axis) for $\Deltasd\taue=2,10,20$ and $\gamma_1=\gamma_2=1$.
 The contribution  $\forcecoef^{(1)}$ is shown by a dashed line.
 The magnon number per cite $n$, plotted on the right axis, indicates that the temperature regime is dilute magnon regime.
\label{FIGmagforce}}
\end{figure}

From Eqs. (\ref{f1result}) (\ref{f2result}), the total force on magnon lowest order in the $sd$ exchange interaction,
$\force_{\rm m}\equiv \force^{(1)}+\force^{(2)} =\forcecoef_{\rm m} eE $, where the coefficient $\forcecoef_{\rm m} \equiv \overline{\forcecoef}^{(1)}+\overline{\forcecoef}^{(2)} $ is
\begin{align}
 \forcecoef_{\rm m} = &
\gamma_1  W^{(1)}(T)+\gamma_2 (\Deltasd \taue)^2 W^{(2)}(T)
\label{totalf}
\end{align}
where
\begin{align}
\gamma_1 &\equiv
 \frac{3}{8S } \gamma_f   \overline{\nel}a^3
\nnr
\gamma_2 &\equiv   \frac{\pi}{3S^2} \gamma_f\ef  \overline{\dos}(1+\tfrac{P_\nu}{4}) \nnr
\gamma_f & \equiv \frac{1}{ma^2 \ef} \label{gammasdef}
\end{align}
where the magnon weight factors are
\begin{align}
W^{(1)}(T)  = &  \frac{a^5}{2\pi^2} \int_0^\infty  dp  p^4 \frac{1}{e^{2\beta \omega_p}-1}
\nnr
W^{(2)}(T) = &   \frac{a^8}{(2\pi^2)^2} \int_0^\infty  dp \int_0^\infty  dp'  p^2 (p')^2(p^2+(p')^2)
  \frac{1}{e^{\beta \omega_p}-1}  \frac{1}{1-e^{-\beta \omega_{p'}}}
 \frac{1}{e^{\beta (\omega_{p'}-\omega_p)}+1}
\end{align}
Considering strong spin polarization in 3$d$ ferromagnets, we may approximate
$\gamma_1\sim\gamma_2$, and then the magnon scattering contribution has a larger coefficient by a factor of $(\Deltasd\taue)^2$ compared to the emission/absorption contribution.
Considering temperatures higher than the magnon gap ($\kb T\gg KS$)
the weight factors in three dimensions are
($x\equiv \beta\omega_p$)
\begin{align}
W^{(1)}(T)  = &
I_f^{(1)} \lt(\frac{\kb T}{\tilde{J}}\rt)^{\frac{5}{2}}
\nnr
W^{(2)}(T) = &
I_f^{(2)} \lt(\frac{\kb T}{\tilde{J}}\rt)^{4}
\end{align}
where $\tilde{J}\equiv JS/a^2$ and
\begin{align}
I_f^{(1)}  &\equiv  \frac{1}{4\pi^2}
  \int_0^\infty  dx  \frac{x^\frac{3}{2}}{e^{2x}-1} = 7.985\times 10^{-3} \nnr
 I_f^{(2)} &\equiv  \frac{1}{(4\pi^2)^2}\int \int_0^\infty  dx  dx'
\frac{x^{\frac{1}{2}} (x')^{\frac{1}{2}}(x+x')}{(e^{x}-1)(1-e^{-x'})(e^{x'-x}+1)}
 = 1.069\times 10^{-2}
\end{align}
The total force coefficient $\forcecoef_{\rm m}$ is plotted as function of temperature in Fig. \ref{FIGmagforce}.
The temperature is normalized by $\tilde{J}$, which is related to the mean-field ferromagnetic transition temperature in three dimensions as
$\kb T_{\rm c}=2(S+1)\tilde{J}$ \cite{Kittel96}.
The temperature regime in Fig. \ref{FIGmagforce} thus corresponds to low temperature  ($T\lesssim T_{\rm c}/4$ for $S=1$).
This is confirmed from the plot of the magnon number per cite (Fig. \ref{FIGmagforce}),
\begin{align}
 n_{\rm m}=a^3\int \frac{p^2dp}{2\pi^2}\frac{1}{e^{\beta\omega_p}-1}
 =I_n \lt(\frac{\kb T}{\tilde{J}}\rt)^{\frac{3}{2}}
\end{align}
where
\begin{align}
 I_n \equiv \frac{1}{4\pi^2}
  \int_0^\infty  dx  \frac{ \sqrt{x}}{e^{x}-1} =0.05864
\end{align}

As is seen in Fig. \ref{FIGmagforce}, the scattering contribution ($\force^{(2)}$) dominates in the wide temperature region, as a result of large factor $(\Deltasd \taue)^2 $.
The crossover temperature from emission/absorption to  scattering regime is
$\kb T_{\rm ea-sc}=\tilde{J} \lt(\frac{\gamma_1 I^{(1)}}{\gamma_2 I^{(2)}} \rt)^{\frac{2}{3}} (\Deltasd\taue)^{-\frac{4}{3}}$.
The crossover temperatures shall be discussed in Sec. \ref{SECcrossover}.

The force on magnon resulting in the $sd$ exchange interaction means that the opposite force acts on electrons.
It reads,
\begin{align}
 {\bm \force}_e &\equiv - ({\bm \force}^{(1)}+{\bm \force}^{(2)}) = -e  {\forcecoef}_{\rm m}\Ev
\end{align}
The coefficient $\forcecoef_{\rm m}$ thus
corresponds to a reduction of the applied electric field as $\Ev\ra (1-\forcecoef_{\rm m})\Ev$. The effect represents a resistance due to magnon scattering and emission.
The electric current taking account magnons is $\jv=\sigmab (1-\forcecoef_{\rm m})\Ev$ ($\sigmab$ is the Boltzmann conductivity).
The magnon contribution to the resistivity in the case of $\forcecoef_{\rm m}\ll 1$ is therefore
\begin{align}
 \delta \rho_{\rm m} &= \rho_0 \forcecoef_{\rm m}
\end{align}
where $\rho_0=1/\sigmab$.
The force on magnon is therefore directly  accessible by the resistivity measurement.

\subsection{Magnon velocity \label{SECmagvelfromF}}
\begin{figure}
 \includegraphics[width=0.8\hsize]{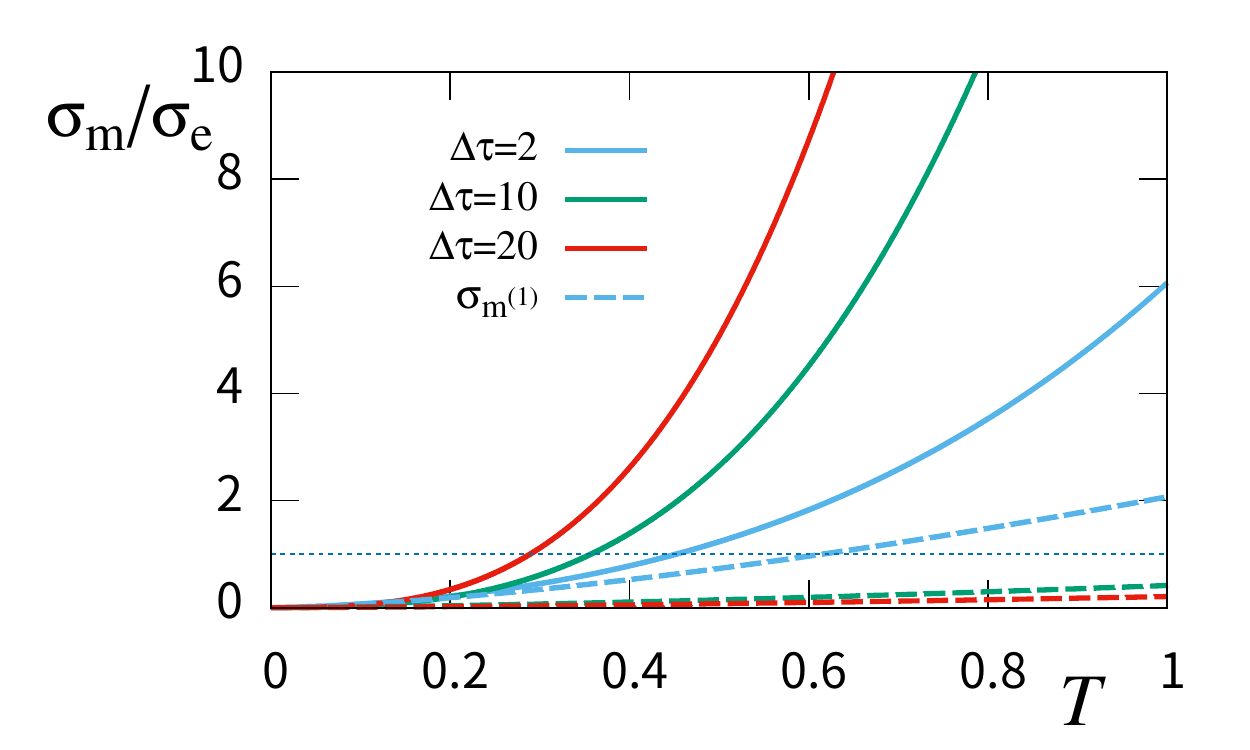}
 \caption{ The ratio $ \frac{\sigma_{\rm m}}{\sigmab} $ of magnon and electron conductivity
 plotted for $\tilde{T}=\kb T/\tilde{J}$ for $\Deltasd\taue=2,10,20$.
 Solid and dashed lines corresponds to the total magnon conductivity and the contribution from the emission/absorption ($\sigma_{\rm m}^{(1)}$), respectively.
 Parameters used are $\gamma_1=\gamma_2=1$ and $\Deltasd=\ef$, $I_v^{(2)}=0.01$ and $\alpha_{\rm G}=0.01$.
 For good metal ($\ef\tau\gtrsim10$), the magnon conductivity is larger than the electric conductivity even at low temperature of $\tilde{T}\gtrsim 0.5$.
 The emission/absorption contribution is smaller than the scattering contribution in this temperature range as a result of a relative suppression factor of $(\Deltasd\taue)^{-2}$.
\label{FIGmagconductivity}}
\end{figure}

When a force $\force_{\rm m}$ acts on a magnon, the magnon is driven at a velocity
$\force_{\rm m}  \frac{\tau_{\rm m}}{m_{\rm m}}$, where $\tau_{\rm m}$ and $m_{\rm m}$ are the magnon lifetime and mass. The lifetime is written in terms of the Gilbert damping parameter $\alpha_{\rm G}$ and the frequency of the magnon $\omega$ as
$1/\tau_{\rm m} =\alpha_{\rm G} \omega$ and magnon mass in the present case is
$m_{\rm m}=(2JS)^{-1}$.
The correctness of the above argument is supported by a direct linear response calculation of the magnon velocity (Appendix \ref{APPmagnonvelociy}).
From our results of the forces, Eq. (\ref{totalf}), the magnon current induced by each force reads $\jv_{\rm m}^{(1)}=\sigma_{\rm m}^{(1)}e\Ev$ and $\jv_{\rm m}^{(2)}=\sigma_{\rm m}^{(2)}e\Ev$, where
\begin{align}
\sigma_{\rm m}^{(1)}
 = &
 \frac{2\gamma_1}{\alpha_{\rm G}a} \lt(\frac{\kb T}{\tilde{J}}\rt)^{\frac{3}{2}} I_v^{(1)}
  \nnr
\sigma_{\rm m}^{(2)}
 = &
 \frac{2\gamma_2}{\alpha_{\rm G}a} (\Deltasd \taue)^2
  \lt(\frac{\kb T}{\tilde{J}}\rt)^{3} I_v^{(2)}
  \label{magnonconductivity}
\end{align}
correspond to conductivity for magnons (without the factor of $e^2$),
with magnon integrals
\begin{align}
I_v^{(1)}  &\equiv  \frac{1}{4\pi^2}
  \int_0^\infty  dx  \frac{x^\frac{1}{2}}{e^{2x}-1} = 2.073\times 10^{-2} \nnr
\nnr
I_v^{(2)}  &\equiv
  \frac{1}{(4\pi^2)^2} \int_{x_0}^\infty  dx \int_0^\infty dx'
\frac{x^{-\frac{1}{2}} (x')^{\frac{1}{2}}(x+x')}{(e^{x}-1)(1-e^{-x'})(e^{x'-x}+1)}
\end{align}.
 The expressions for $ I_v^{(2)} $ diverges at low energy due to insufficient  phenomenological treatment of magnon lifetime in terms of the Gilbert damping in the low frequency limit.
 Here we avoid the problem by introducing a low energy cutoff $\omega_0$ ($x_0=\omega_0^2/(\kb T)$).
$I_v^{(2)}$ is of the order of 0.01 for $x_0\gtrsim 0.02$ and logarithmically diverges as $x_0\ra 0$.
The electric conductivity  in the same approximation
($1/(2ma^2)\sim \ef$, $\kf a\sim1$, $\overline{\dos}\sim 1/\ef$) is
$\sigmab/e^2 \simeq \frac{1}{a}\ef\taue$.
The ratio of the magnon conductivity to the electron one thus is
\begin{align}
\mu_{\rm m/e}
= \frac{\sigma_{\rm m}}{\sigmab}
  &\sim \frac{2}{\alpha_{\rm G}\ef\taue}
  \lt[ \gamma_1 I_v^{(1)} \tilde{T}^{\frac{3}{2}}
      +\gamma_2 I_v^{(2)} ({\Deltasd \taue})^2 \tilde{T}^{{3}}  \rt]
  \label{sigmamsigmaeratio}
\end{align}
where $\tilde{T}\equiv \frac{\kb T}{\tilde{J}}$.
As seen from Fig. \ref{FIGmagconductivity}, in good metals with large $\Deltasd\taue$,  magnon conductivity is larger than the electric conductivity even for a low temperature   (e.g., $\tilde{T}\gtrsim 0.4$   for $\ef\taue=20$).
The result is qualitatively consistent with seminal work of Grannemann and Berger \cite{Grannemann76}, where it was argued that average drift velocity of magnon is 2-3 times larger than that of electron in Ni$_{66}$Cu$_{34}$ and Ni$_{69}$Fe$_{31}$.
Our result for $\sigma_{\rm m}^{(1)}$ is consistent with previous analysis \cite{Miura12}, where energy current driven by an electric field was microscopically calculated taking account of the magnon emission/absorption  ($f^{(1)}$) .
Our result, however, indicates that more efficient magnon-drag effect occurs due to the magnon scattering ($f^{(2)}$) in strongly spin-polarized good metals.
The reason is that the scattering contribution is a response of the electron charge sector (spin summed), while the emission/absorption is a spin response containing electron propagators with opposite spins, resulting in a relative suppression factor of
$(\Deltasd\taue)^{-2}$.
As the electron elastic lifetime is long, $\ef\taue\gg1$, except for extremely dirty metals, the enhancement factor of  $(\Deltasd\taue)^{2}$ for scattering contribution makes the contribution larger even for temperatures of
$\kb T \gtrsim (\Deltasd\taue)^{-\frac{4}{3}}\tilde{J}$
(See Sec. \ref{SECcrossover}).

The expression for the emission/absorption, $\sigma_{\rm m}^{(1)}$, of Eq. (\ref{magnonconductivity}) is consistent with previous analysis indicating
$\sigma_{\rm m}^{(1)}\propto \frac{1}{\alpha_{\rm G} \taue}T^{\frac{3}{2}}$ \cite{Grannemann76,Miura12}, although
the result of Ref. \cite{Miura12} is proportional to electron spin polarization,
$\sigma_{\rm m}^{(1)}\propto \frac{P}{\alpha_{\rm G} \taue}T^{\frac{3}{2}}$, probably due to an insufficient treatment of magnon and  hole contributions in Ref. \cite{Miura12}.

The result for
$\sigma_{\rm m}^{(2)}$ suggests that the magnon damping effect for low energy magnons are critical for estimation of the magnon conductivity.
Further theoretical and experimental investigations are expected in this direction.

\subsection{Magnon spin-transfer effect}
\begin{figure}
 \includegraphics[width=0.8\hsize]{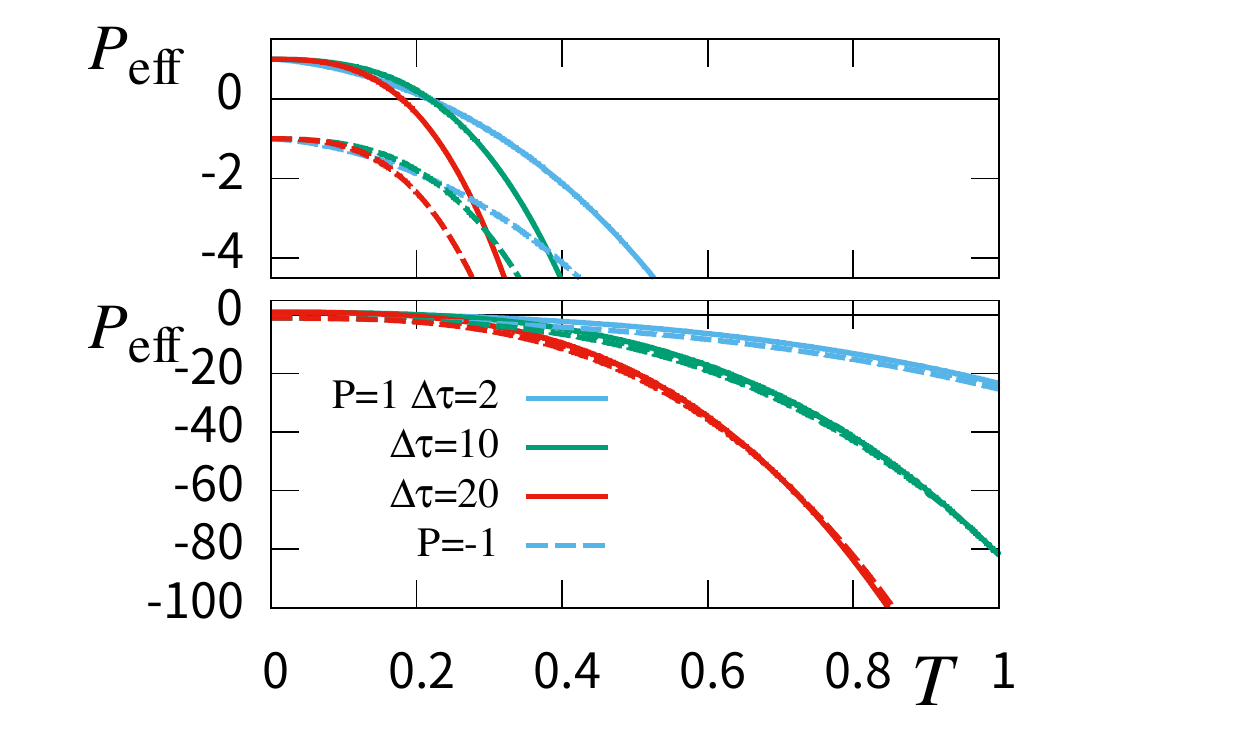}
 \caption{ The effective spin-transfer efficiency $P_{\rm eff}$ plotted for $\tilde{T}=\kb T/\tilde{J}$ for $\Deltasd\taue=2,10,20$ and $P=1$ (solid lines) and $P=-1$ (dashed lines). Parameters used are $\gamma_1=\gamma_2=1$ and $\Deltasd=\ef$, $I_v^{(2)}=0.01$ and $\alpha_{\rm G}=0.01$.
\label{FIGPeff}}
\end{figure}

The magnon current can be estimated by observing magnetization dynamics induced by the magnon spin-transfer effect.
In the case of electron spin-transfer effect, the flow of magnetization structures is at the velocity of
\begin{align}
v_{\rm st,e} &= \frac{a^3 P}{2eS}j
\end{align}
in the direction of spin current $Pj$ ($P=\frac{n_\uparrow -n_\downarrow}{n_\uparrow +n_\downarrow}$)
in the adiabatic limit \cite{TKS_PR08}.
The current is written in terms of the current in the low temperature limit (i.e., without magnons) $j^{(0)}$ as
$j=j^{(0)}(1-\forcecoef_{\rm m})$, where $\forcecoef_{\rm m}$ represents the resistivity effect due to magnons.
Assuming the adiabatic limit for magnons, the magnon spin transfer effect drives magnetization structures at the velocity of
\begin{align}
v_{\rm st,m} &= - \frac{a^3}{S}j_{\rm m}
\end{align}
in the opposite direction to the magnon current.
Considering the two mechanisms for magnon current, the velocity of the magnetization structure for an applied electric current $\jv$ is
\begin{align}
 v_{\rm st} &= \frac{a^3 }{2eS}P_{\rm eff}j^{(0)}
 \label{stttotal}
\end{align}
where
\begin{align}
 P_{\rm eff} &\equiv
 P\lt[1-\forcecoef_{\rm m} \rt]
 - \frac{4}{\alpha_{\rm G}\ef\taue}
  \lt[ \gamma_1  I_v^{(1)} \tilde{T}^{\frac{3}{2}}
      +\gamma_2  I_v^{(2)} ({\Deltasd \taue})^2 \tilde{T}^{3}  \rt]
   \label{Peff}
\end{align}
represents the effective spin-transfer efficiency including the magnon effects, plotted in Fig. \ref{FIGPeff}.
In the weak damping regime $\alpha_{\rm G}\lesssim (\ef\taue)^{-1}$, the contribution of $\forcecoef_{\rm m}$ (magnon resistivity) is negligible compared to the conductivity correction (the last square bracket) in the temperature regime in Fig. \ref{FIGPeff}.
Crossover from electron-dominated to the magnon-dominated regime occurs at
$\kb T/\tilde{J}\lesssim 0.2$ in the present case, with a significant negative enhancement in the high temperature regime.
The magnon spin-transfer effect is correlated with the behavior of magnon contribution to the resistance, that is proportional to the force plotted in Fig. \ref{FIGmagforce}.  Identification of magnon drag effects would be carried out by careful analyses of temperature dependence of the experimental data.

\subsection{Crossover temperatures \label{SECcrossover}}

Let us look into the crossover temperatures based on the results Eqs. (\ref{totalf}) (\ref{Peff}).
We consider the case $\gamma_1\sim\gamma_2\sim1$, $\Deltasd/\ef\sim 1$ and neglect the contribution from
$\forcecoef_{\rm m}$ in $P_{\rm eff}$ for simplicity.
As for the magnons, emission/absorption effect is dominant at low temperature, and the crossover to the scattering-dominated temperature is read from Eq. (\ref{totalf}) as
\begin{align}
 T_{{\rm ea-sc}} \simeq (\Deltasd\taue)^{-\frac{4}{3}} \tilde{J}/\kb
  \label{Teasc}
\end{align}
The crossover would be seen in the  magnon resistivity, assuming that drag force is not directly observable.

\begin{figure}
 \includegraphics[width=0.8\hsize]{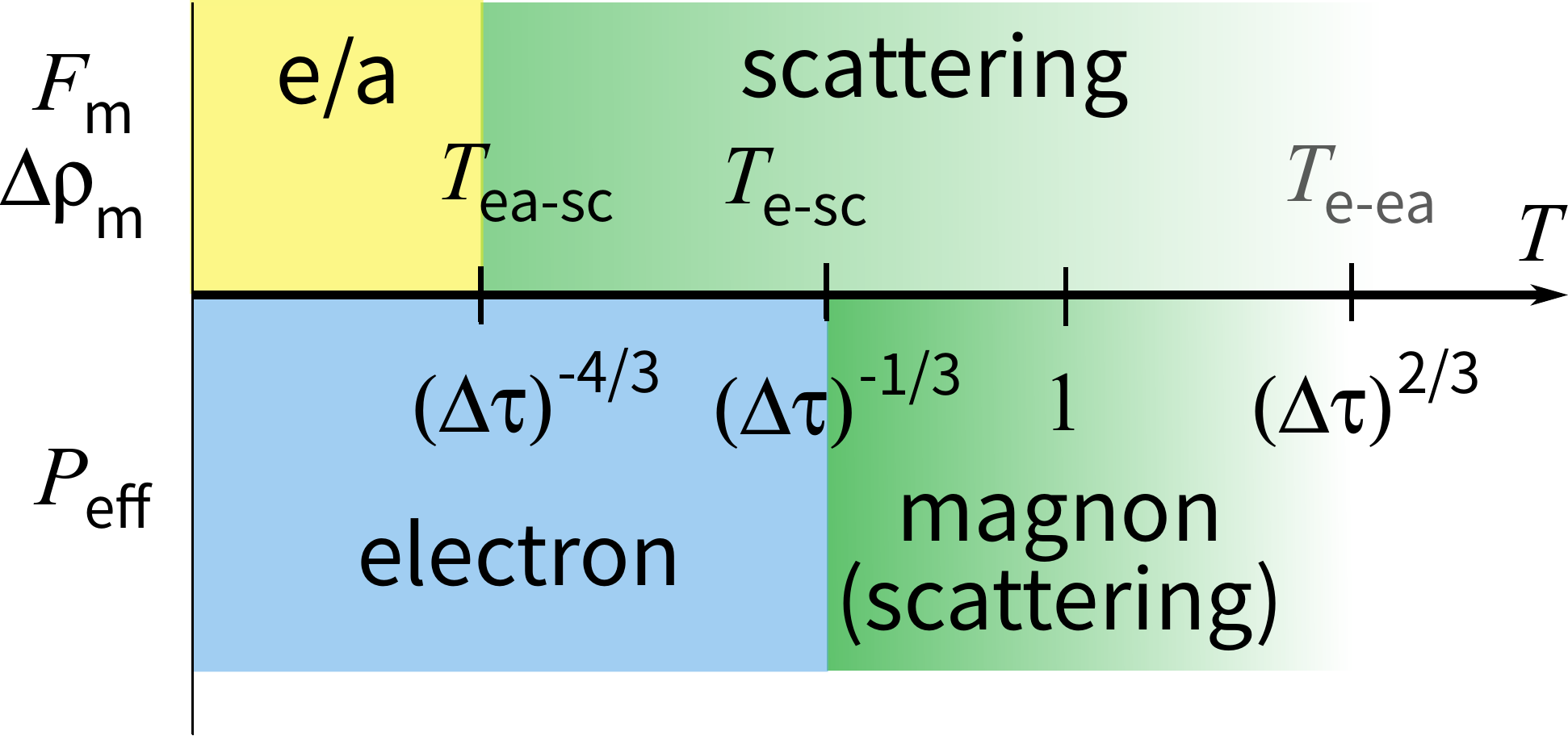}
 \caption{ Schematic figure showing the crossover temperatures.
 In the magnon force $F_{\rm m}$ and magnon contribution to the electric resistivity ($\Delta\rho_{\rm m}$), crossover from magnon emission/absorption (e/a) to scattering is at $T_{\rm ea-sc}$.
 The effective spin polarization (spin transfer efficiency) $P_{\rm eff}$ is dominated by the electron contribution at low temperature while a crossover to magnon spin-transfer dominated regime driven by magnon scattering occurs at $T_{\rm e-sc}$.
 The order of magnitude of $P_{\rm eff}$ in terms of powers of $\Delta\tau$ is shown. The magnon emission/absorption process, relevant above $T_{\rm e-ea}$,d would not be dominant in $P_{\rm eff}$ in the dilute magnon regime $\kb T/\tilde{J} \lesssim 1$ unless in very dirty metal with $\Delta\tau\sim O(1)$.
\label{FIGcrossover}}
\end{figure}

The spin-transfer efficiency, $P_{\rm eff}$, at zero temperature reduces to the electron origin, $P$.
The magnon emission/absorption contribution becomes larger than the electron contribution
above
\begin{align}
 T_{{\rm e-ea}} \simeq (12\times \alpha_{\rm G})^{\frac{2}{3}} (\Deltasd\taue)^{\frac{2}{3}} \tilde{J}/\kb
  \label{Teea}
\end{align}
which corresponds to high temperature $ T_{{\rm e-ea}} \gtrsim \tilde{J}/\kb$ unless in an extremely low damping materials with $\alpha_{\rm G} \ll 0.08\times (\Deltasd\taue)^{-1}$.
In contrast, magnon scattering effect overcomes the electron contribution at lower temperature of
\begin{align}
 T_{{\rm e-sc}} \simeq (25\times \alpha_{\rm G})^{\frac{1}{3}} (\Deltasd\taue)^{-\frac{1}{3}} \tilde{J}/\kb
  \label{Tesc}
\end{align}
for $I_v^{(2)}=0.01$, meaning that the magnon emission/absorption regime emerges only in very dirty metals with $\Deltasd\taue\sim O(1)$ (Fig. \ref{FIGPeff}).
The crossover behavior is summarized in Fig. \ref{FIGcrossover}.

\section{Summary}
We have calculated the force between the magnon and conduction electron when an electric field is applied to a ferromagnetic metal based on a microscopic approach.
The force due to magnon emission/absorption and scattering were considered and the latter turned out to dominate  in a wide temperature regime in good metals with long elastic mean free path.
The magnon contribution to the resistivity and total spin-transfer efficiency were discussed.

\acknowledgements
This study was supported by
a Grant-in-Aid for Scientific Research (B) (No. 21H01034) from the Japan Society for the Promotion of Science.

\appendix
\section{Properties of magnon-electron composite propagators \label{SECAPPmepropagator}}

\begin{figure}
 \includegraphics[width=0.3\hsize]{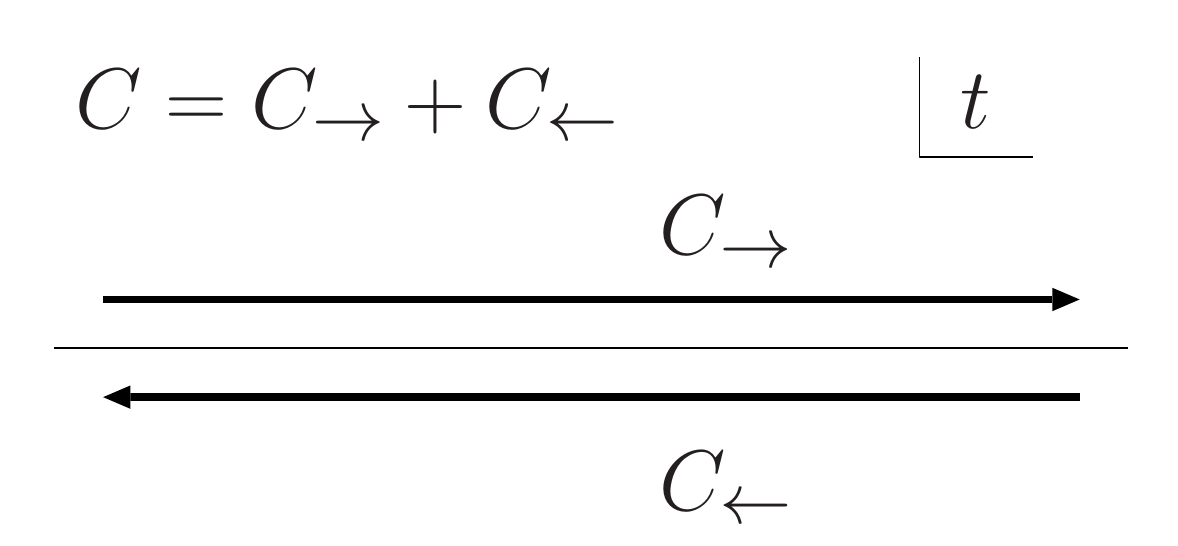}
 \caption{The time contour the path-ordered Green's functions are defined. The upper (lower) path is $C_\rightarrow$ ($C_\leftarrow$).
\label{FIGC}}
\end{figure}

The path-ordered Green's functions, defined on a time contour $C$ (Fig. \ref{FIGC}), are written in terms of real-time Green's functions as
\begin{align}
G^{--}(t,t') &\equiv G(t\in C_\rightarrow,t'\in C_\rightarrow)
= \theta(t-t')G^>(t,t')+\theta(t'-t)G^<(t,t') =G^{\rm t}(t,t')  \nnr
G^{-+}(t,t') &\equiv G(t\in C_\rightarrow,t'\in C_\leftarrow)=G^<(t,t') \nnr
G^{+-}(t,t') &\equiv G(t\in C_\leftarrow,t'\in C_\rightarrow)=G^{>}(t,t')  \nnr
G^{++}(t,t') &\equiv G(t\in C_\leftarrow,t'\in C_\leftarrow)
= \theta(t-t')G^<(t,t')+\theta(t'-t)G^>(t,t')=G^{\rm \bar{t}}(t,t')
\label{GFs}
\end{align}
where the time on the path $C_\rightarrow$ and $C_\leftarrow$ are denoted as $-$ and $+$, respectively, and $G^{\rm t}$ and $G^{\rm \bar{t}}$ are the time-ordered and anti-time-ordered Green's functions.
These expressions are direct consequence of the definition of path ordering on $C$.
A straightforward relation derived from Eq. (\ref{GFs}) is
\begin{align}
G^{--}(t,t') + G^{++}(t,t')
= G^{-+}(t,t') + G^{+-}(t,t')
= G^{\rm K}(t,t')
\equiv [G^<+G^>](t,t')
\label{GFrelation1}
\end{align}
Noting the relations
\begin{align}
G^{\rm t} &= G^<+G^\ret = G^>+G^\adv \nnr
G^{\rm \bar{t}} &= G^<-G^\adv = G^> -G^\ret,
\label{GtGtb}
\end{align}
which read
$G^\adv=G^{--}-G^{+-}$ and $G^\ret=G^{+-}-G^{++}$,
the definition (\ref{GFs}) is summarized as
\begin{align}
 G^{\alpha\beta}(t,t')
 &= \frac{1}{2} [ G^{\rm K}(t,t')-\alpha G^\adv(t,t') -\beta G^\ret(t,t') ]
 \label{realtimedecompose}
\end{align}
where $\alpha,\beta=\pm$ are labels representing the path for the time,
$G^\adv$ and $G^\ret$ are the advanced and retarded Green's functions.
\subsection{Magnon electron pair propagator \label{APPmagelecprop}}
An advantage to define the pair propagators defined in Eq. (\ref{Pidefs})  is that the definition has the same structure with respect to time as  the single particle case (\ref{GFs}) and thus the propagators
satisfy the same relation as Eqs. (\ref{GFrelation1})(\ref{GtGtb}).
They therefore satisfy in parallel to Eq. (\ref{realtimedecompose})
\begin{align}
  \Pi_{k}^{(\pm),(k\mp p,p),{\alpha\beta}}
 &= \frac{1}{2}\lt(\Pi_{k}^{(\pm),(k\mp p,p),{\rm K}}-\alpha \Pi_{k}^{(\pm),(k\mp p,p),{\adv}} -\beta \Pi_{k}^{(\pm),(k\mp p,p),{\ret}}\rt)
 \label{realtimedecomposePi}
\end{align}
where
\begin{align}
  \Pi_{k}^{(+),(k-p,p),\mu}(t,t')
     &= i[g_{k-p}^\mu(t,t') d_p^>(t,t')+g_{k-p}^<(t,t') d_p^\mu(t,t')] \nnr
  \Pi_{k}^{(+),(k-p,p),\nu}(t,t')
     &= i[g_{k-p}^\nu(t,t') d_p^\nu(t,t')]
     \label{Pi+}
\end{align}
for $\mu=\ret,\adv$ and $\nu=>,<$ (with $\Pi^{\rm K}\equiv \Pi^<+\Pi^>$)
and
\begin{align}
  \Pi_{k}^{(-),(k+p,p),\adv}(t,t')
     &= i[g_{k+p}^\adv(t,t') d_p^<(t',t)+g_{k+p}^<(t,t') d_p^\ret(t',t)] \nnr
  \Pi_{k}^{(-),(k+p,p),\ret}(t,t')
     &= i[g_{k+p}^\ret(t,t') d_p^<(t',t)+g_{k+p}^<(t,t') d_p^\adv(t',t)] \nnr
  \Pi_{k}^{(-),(k+p,p),<}(t,t')
     &= i[g_{k+p}^<(t,t') d_p^>(t',t)] \nnr
  \Pi_{k}^{(-),(k+p,p),>}(t,t')
     &= i[g_{k+p}^>(t,t') d_p^<(t',t)] \label{Pi-}
\end{align}
This fact indicates that the pair propagator indeed behaves as a propagator of a composite particle.
In the frequency representation,
\begin{align}
  \Pi_{k}^{(+),(k+p,p),>}(\omega)
  &\equiv \int dt e^{i\omega t}
  \Pi_{k}^{(+),(k+p,p),>}(t)
     = 2\pi i \sum_{\omega_1} g_{k+p}^>(\omega_1) d_p^>(-\omega_1-\omega)\nnr
  \Pi_{k}^{(-),(k+p,p),>}(\omega)
  &\equiv \int dt e^{i\omega t}
  \Pi_{k}^{(-),(k+p,p),>}(t)
     = 2\pi i \sum_{\omega_1} g_{k+p}^>(\omega_1) d_p^<(\omega_1-\omega)
\end{align}
Moreover, the pair propagator satisfy the the same relation as single particle Green's function without dynamic interaction, i.e.,
\begin{align}
  \Pi_{k}^{(\pm),(k\mp p,p),<}(\omega)
     &= f(\omega)[
  \Pi_{k}^{(\pm),(k\mp p,p),\adv}(\omega)-
  \Pi_{k}^{(\pm),(k\mp p,p),\ret}(\omega)]
  \label{Pistaticrelation}
\end{align}
which is useful to extract the low energy contributions.

For the $\omega=0$ component of the advanced pair propagator in Eq. (\ref{fijeq1}), we have, using Eqs. (\ref{Pi+})(\ref{Pi-}),
\begin{align}
 \Pi_{k\sigma}^{(+),(k-p,p),\adv}
 &=
i \sum_\nu [g_{k-p,\nu,\sigma}^\adv d^>_{p,-\nu}+g_{k-p,\nu,\sigma}^< d^\adv_{p,-\nu}]
\end{align}
where $\nu$ is the frequency for electron Green's function.
The frequency of magnon is $-\nu$ as the total frequency of the pair propagator
$\pi^{(+)}$ is zero
and
\begin{align}
 \Pi_{k\sigma}^{(-),(k+p,p),\adv}
 &=
i \sum_\nu [g_{k+p,\nu,\sigma}^\adv d^<_{p,\nu}+g_{k+p,\nu,\sigma}^< d^\ret_{p,\nu}]
\end{align}
where magnon frequency is equal to the electron one for $\Pi^{(-)}$.
We thus obtain
\begin{align}
 \Im \Pi_{k\sigma}^{(+),(k-p,p),\adv}
 &=
4 \sum_\nu n_\nu(1-f_\nu) \Im [g_{k-p,\nu,\sigma}^\adv] \Im[d^\adv_{p,-\nu}]]
\nnr
 \Im \Pi_{k\sigma}^{(-),(k+p,p),\adv}
 &=
- 4 \sum_\nu n_\nu(1-f_\nu) \Im [g_{k+p,\nu,\sigma}^\adv] \Im[d^\adv_{p,\nu}]]
\label{ImPi2result}
\end{align}
where we used
$f_{-\nu}=1-f_\nu$ and $n_\nu+f_\nu=2 n_\nu(1-f_\nu)=2(1+n_\nu)f_{\nu}$

\subsection{Two-magnon electron composite propagator \label{APPtwomagelecprop}}
Two-magnon electron composite propagator defined by Eq. (\ref{twomagelecpropagator}) has the same mathematical structure as the magnon-electron pair propagator and single particle Green's functions.
Namely, by definitions
($\pm$ denotes the time contour $C_\ra$ and $C_\la$)
\begin{align}
\Pi^{(2),k',p,p'}_{k\sigma}(t,t')
  & \equiv
  g_{\kv'\sigma}(t,t') d_{\pv}(t,t') d_{\pv'}(t',t) \nnr
\Pi^{(2),k',p,p'(--)}_{k\sigma}(t,t')
  & =
  \theta(t'-t) g_{\kv'\sigma}^<(t,t') d_{\pv}^<(t,t') d_{\pv'}^>(t',t)
+ \theta(t-t') g_{\kv'\sigma}^>(t,t') d_{\pv}^>(t,t') d_{\pv'}^<(t',t)
\nnr
\Pi^{(2),k',p,p'(++)}_{k\sigma}(t,t')
  & =
  \theta(t-t') g_{\kv'\sigma}^<(t,t') d_{\pv}^<(t,t') d_{\pv'}^>(t',t)
+ \theta(t'-t) g_{\kv'\sigma}^>(t,t') d_{\pv}^>(t,t') d_{\pv'}^<(t',t)
\end{align}
we have
\begin{align}
\Pi^{(2),k',p,p'(--)}_{k\sigma}(t,t')
+\Pi^{(2),k',p,p'(++)}_{k\sigma}(t,t')
&=
\Pi^{(2),k',p,p'(-+)}_{k\sigma}(t,t')
+\Pi^{(2),k',p,p'(+-)}_{k\sigma}(t,t')
\end{align}
and this relation allows us to write ($\alpha, \beta=\pm$)
\begin{align}
\Pi^{(2),k',p,p'(\alpha\beta)}_{k\sigma}(t,t')
  & = \frac{1}{2} \lt[
\Pi^{(2),k',p,p',K}_{k\sigma}(t,t')
-\alpha \Pi^{(2),k',p,p',\adv}_{k\sigma}(t,t')
-\beta \Pi^{(2),k',p,p',\ret}_{k\sigma}(t,t') \rt]
\label{Funakirelation}
\end{align}
where
\begin{align}
\Pi^{(2),k',p,p',K}_{k\sigma}
 &\equiv (\Pi^{(2),k',p,p'(++)}_{k\sigma}+\Pi^{(2),k',p,p'(--)}_{k\sigma}) \nnr
\Pi^{(2),k',p,p',\adv}_{k\sigma}
 &\equiv (\Pi^{(2),k',p,p'(-+)}_{k\sigma}-\Pi^{(2),k',p,p'(++)}_{k\sigma})
 =(\Pi^{(2),k',p,p'(--)}_{k\sigma}-\Pi^{(2),k',p,p'(+-)}_{k\sigma}) \nnr
\Pi^{(2),k',p,p',\ret}_{k\sigma}
 &\equiv (\Pi^{(2),k',p,p'(+-)}_{k\sigma}-\Pi^{(2),k',p,p'(++)}_{k\sigma})
 =(\Pi^{(2),k',p,p'(--)}_{k\sigma}-\Pi^{(2),k',p,p'(-+)}_{k\sigma})
\end{align}
Note that here retarded, advanced components are not in the original sense for the single particle Green's functions, written in terms of (anti)commutators of field operators.
Nevertheless, the relation (\ref{Funakirelation}) indicates that multi particle propagators behaves mathematically the same as single particle Green's functions, and it simplifies the calculation greatly.

Using explicit expressions
\begin{align}
\Pi^{(2),k',p,p'(--)}_{k\sigma}(t,t')
  & =
  g_{\kv'\sigma}^<(t,t') d_{\pv}^<(t,t') d_{\pv'}^<(t',t)
+  g_{\kv'\sigma}^\ret(t,t') d_{\pv}^<(t,t') d_{\pv'}^<(t',t)
\nnr &
+  g_{\kv'\sigma}^<(t,t') d_{\pv}^<(t,t') d_{\pv'}^\ret(t',t)
+  g_{\kv'\sigma}^>(t,t') d_{\pv}^\ret(t,t') d_{\pv'}^<(t',t)
  \nnr
\Pi^{(2),k',p,p'(-+)}_{k\sigma}(t,t')
  & =
  g_{\kv'\sigma}^<(t,t') d_{\pv}^<(t,t') d_{\pv'}^>(t',t)
  \nnr
\Pi^{(2),k',p,p'(+-)}_{k\sigma}(t,t')
  & =
  g_{\kv'\sigma}^>(t,t') d_{\pv}^>(t,t') d_{\pv'}^<(t',t)
\end{align}
we obtain
\begin{align}
\Pi^{(2),k',p,p',K}_{k\sigma}(t,t')
 & =
 \lt[ g_{\kv'\sigma}^<(t,t') d_{\pv}^<(t,t') d_{\pv'}^>(t',t)
 +  g_{\kv'\sigma}^>(t,t') d_{\pv}^>(t,t') d_{\pv'}^<(t',t) \rt]
  \nnr
  \Pi^{(2),k',p,p',\adv}_{k\sigma}(t,t')
 &=
 \lt[ g_{\kv'\sigma}^<(t,t') d_{\pv}^<(t,t') d_{\pv'}^\ret(t',t)
+ g_{\kv'\sigma}^<(t,t') d_{\pv}^\adv(t,t') d_{\pv'}^<(t',t)
+  g_{\kv'\sigma}^\adv(t,t') d_{\pv}^>(t,t') d_{\pv'}^<(t',t) \rt]
 \nnr
\Pi^{(2),k',p,p',\ret}_{k\sigma}(t,t')
 &=
  \lt[ g_{\kv'\sigma}^<(t,t') d_{\pv}^\ret(t,t') d_{\pv'}^<(t',t)
+ g_{\kv'\sigma}^<(t,t') d_{\pv}^<(t,t') d_{\pv'}^\adv(t',t)
+  g_{\kv'\sigma}^\ret(t,t') d_{\pv}^>(t,t') d_{\pv'}^<(t',t) \rt]
\label{Pi2components}
\end{align}
The advanced component can be derived also by use of decompose relation and Eq. (\ref{Pi+}) as
\begin{align}
  \Pi^{(2),k',p,p',\adv}_{k\sigma}(t,t')
 &=
 -i[ \Pi_{k+p}^{(+)(k,p)}(t,t') d_{p'}(t',t)]^\adv \nnr
 &= -i[ \Pi_{k+p}^{(+)(k,p),\adv}(t,t') d^<_{p'}(t',t) +  \Pi_{k+p}^{(+)(k,p),<}(t,t') d^\ret_{p'}(t',t)]
 \nnr
 &=
 ( g_{\kv\sigma}^<(t,t') d_{\pv}^<(t,t') d_{\pv'}^\ret(t',t)
+ g_{\kv\sigma}^<(t,t') d_{\pv}^\adv(t,t') d_{\pv'}^<(t',t)
+  g_{\kv\sigma}^\adv(t,t') d_{\pv}^>(t,t') d_{\pv'}^<(t',t) )
\end{align}
indicating consistency of composite propagator representation.
The advanced component of the composite propagator in the Fourier representation is
\begin{align}
\Pi^{(2),k',p,p',\adv}_{k\sigma}
  &=
  \sum_{\nu\nu'} [
  g_{\kv',\nu'-\nu,\sigma}^\adv d_{\pv\nu}^< d_{\pv'\nu'}^>
 + g_{\kv',\nu'-\nu,\sigma}^> d_{\pv\nu}^\adv d_{\pv'\nu'}^<
 +   g_{\kv',\nu'-\nu,\sigma}^> d_{\pv\nu}^< d_{\pv'\nu'}^\ret ]
 \label{Pi2adv}
\end{align}
Using $\Im[d_{p\nu}^\adv]=\pi\delta(\nu-\omega_p)$, we have
\begin{align}
\Im[\Pi^{(2),k',p,p',\adv}_{k\sigma}]
  &=
 - [ n_{\omega_p}(1+n_{\omega_p'}) f_{\omega_{p'}-{\omega_p}}+n_{\omega_{p'}}(1+n_{\omega_p}) f_{{\omega_p}-{\omega_p'}}]
\Im[g_{\kv',\nu'-\nu,\sigma}^\adv   ]
\label{ImPi2a}
\end{align}

\section{Calculation of electron contribution of Eq. (\ref{electronsumf1}) \label{SECelectronsum}}

\begin{figure}
 \includegraphics[width=0.3\hsize]{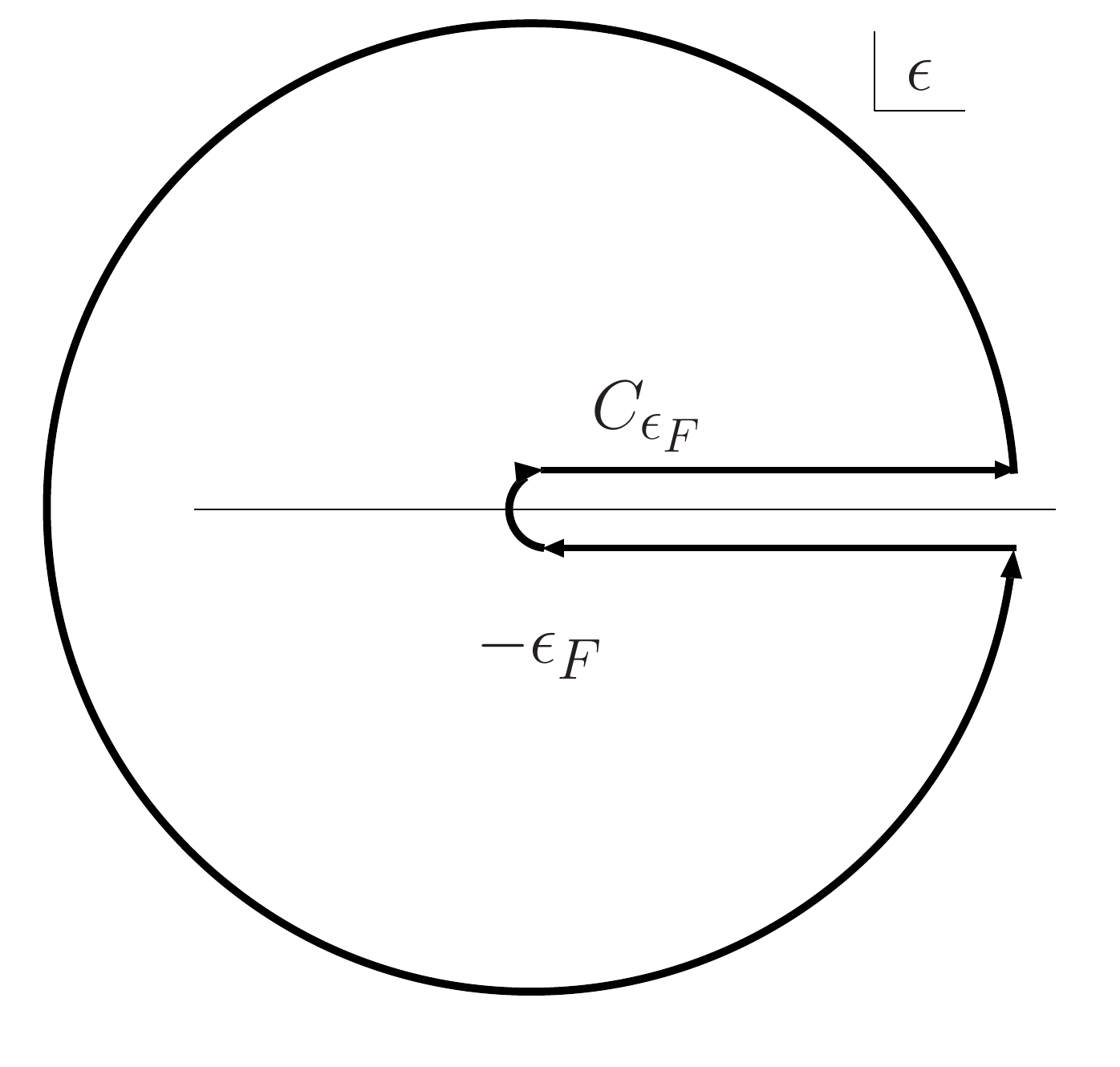}
 \caption{energy contour $C_{\epsilon_F}$  for calculating the $k$-summation of the electron Green's function.
\label{FIGCenergy}}
\end{figure}
The electron contribution, Eq. (\ref{electronsumf1}),
to the emission/absorption force $F^{(1)}$ is calculated here by use of a contour integration. We first write
$I_\sigma \equiv \sum_{\kv}k^2 |g_{k\sigma}^\adv|^2 (g_{k,-\sigma}^\adv)^2$
by use of the energy ($\epsilon\equiv\epsilon_k$) integral as
\begin{align}
 I_\sigma & = \int_{-\ef}^\infty d\epsilon \frac{ \dos(\epsilon)k^2(\epsilon) }{(\epsilon-\sigma \Deltasd)^2+\eta^2} \frac{1}{(\epsilon+\sigma\Deltasd+i\eta)^2}
\end{align}
where $\dos(\epsilon)=\frac{mk(\epsilon)\al^3}{2\pi^2}$ is the electron  density of states, $k(\epsilon)\equiv \sqrt{2m(\epsilon+\ef)}$.
The integration is written as an integration over a contour $C_{\ef}$ avoiding a cut along the real axis (due to $\dos(\epsilon)\sqrt{\epsilon+\ef}$) shown in Fig. \ref{FIGCenergy}as
\begin{align}
 I_\sigma & = \frac{1}{2}\int_{C_{\ef}} d\epsilon \frac{ \dos(\epsilon)k^2(\epsilon) }{(\epsilon-\sigma \Deltasd)^2+\eta^2} \frac{1}{(\epsilon+\sigma\Deltasd+i\eta)^2}
\end{align}
The residues at poles $\epsilon=\sigma\Deltasd\pm i\eta$ and $\epsilon=-\sigma\Deltasd-i\eta$ are calculated, paying attention to the fact that
$\dos(\epsilon-i\eta)=-\dos(\epsilon+i\eta)$ due to the cut, to obtain
\begin{align}
 I_\sigma & = \frac{\pi}{2\eta}\dos_\sigma \frac{1}{4}
 \lt[ \frac{1}{\Deltasd^2} + \frac{1}{(\Deltasd+i\sigma \eta)^2} \rt]
 +i\pi \frac{d}{d\epsilon}\lt[
 \frac{ \dos(\epsilon)k^2(\epsilon) }{\epsilon-\sigma \Deltasd+i\eta} \frac{1}{\epsilon-\sigma\Deltasd-i\eta}
 \rt]_{\epsilon=-\sigma\Deltasd-i\eta}
\end{align}
It turns out that the lowest order contribution in the limit of $\eta/\ef\ll1$ arises from the derivative of $\nu(\epsilon)k^2(\epsilon) $, i.e.,
\begin{align}
 I_\sigma & = -i\pi \frac{3}{8\ef}\dos_{-\sigma} k^2_{-\sigma}
 \frac{1}{\Deltasd(\Deltasd+i\sigma \eta)}
 \end{align}
 We therefore obtain
\begin{align}
  \Im \sum_{\sigma}I_\sigma
 &=
 -\frac{3\pi}{8} \frac{1}{\Deltasd^2 \ef} \sum_{\sigma}\sigma{\dos_{\sigma} k_{\sigma}^2}
 =
 -\frac{9\pi}{8} \frac{m a^3 \nel}{\Deltasd^2 \ef}
\end{align}
where $\nel=\frac{1}{3m\al^3}\sum_{\sigma}\sigma{\dos_{\sigma} k_{\sigma}^2}$ is the total electron density.
\section{Linear response calculation of magnon current \label{APPmagnonvelociy}}
Here we calculate the magnon current driven by an electric field directly diagrammatically to show the correctness of the argument in Sec. \ref{SECmagvelfromF} of the velocity based on the calculation result of the force on magnon.
The calculation of the emission/absorption contribution is essentially the same as the one for the magnon energy current in Ref. \cite{Miura12}, while scattering contribution was not argued there.
Magnon current is
\begin{align}
  \jv_{\rm m} &= -i{JS}  \magnon^\dagger \stackrel{\leftrightarrow}{\nabla} \magnon \nnr
  &=JS(\nabla_{r}-\nabla_{r'}) D^<(\rv,t,\rv',t)|_{\rv=\rv'}
  \label{jmdef}
\end{align}
where $ D^<(\rv,t,\rv',t')=-i\average{\magnon^\dagger(\rv',t')\magnon(\rv,t)}$ is the full Greens function of magnon including interactions (the sign of $-$ is for boson lesser Green's function).
The full Green's function at the linear response is perturbatively expanded including the $sd$ exchange interaction to the second order and the electric field

\subsection{Magnon emission/absorption contribution}

\begin{figure}
 \includegraphics[width=0.5\hsize]{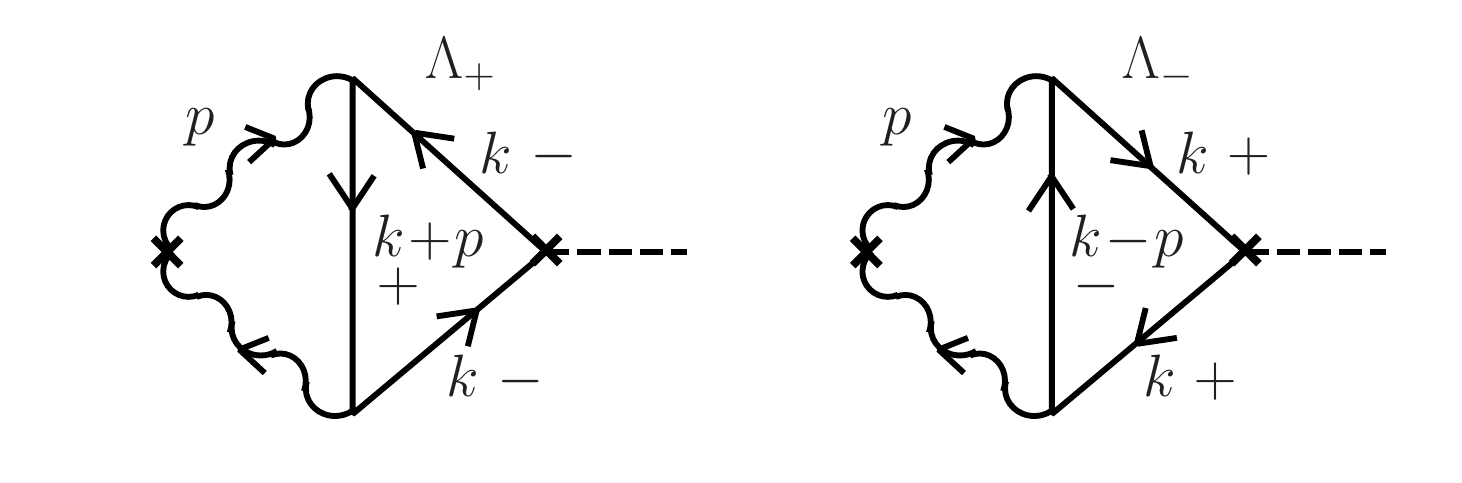}
 \caption{
Feynman diagrams representing the magnon current (left vertex) induced by an electric field (denoted by a dotted line) by the magnon emission/absorption due to the $sd$ exchange interaction. Wavy and solid lines denote magnon and electron propagators, respectively. The sign $\pm$ denotes electron spin.
\label{FIGmagj}}
\end{figure}

Here magnon current $j_{\rm m}^{(1)}$ due to the linear magnon coupling to electrons arising from the $sd$ exchange interaction is calculated.
The corresponding contribution to the Green's function defined on the time contour,$D^{(1)}(\rv,t,\rv',t')$, is (diagrams shown in Fig. \ref{FIGmagj})
\begin{align}
 D^{(1)}(\rv,t,\rv',t')
  &=
  i\frac{e}{m}\frac{\Deltasd^2}{2S} \int_C dt_1 \int_C dt_2 \int_C dt_3
  \sum_{r_1r_2r_3} d(\rv_2,t_2,\rv',t') d(\rv,t,\rv_1,t_1)  A_j(t_3) \nnr
 & \times
 \tr [\sigma_+g({\rv_1,t_1},\rv_2,t_2) \sigma_- g(\rv_2,t_2,\rv_3,t_3) \hat{p}_jg(\rv_3,t_3,\rv_1,t_1)
 \nnr&
 +\sigma_+g({\rv_1,t_1},\rv_3,t_3) \hat{p}_j g(\rv_3,t_3,\rv_2,t_2) \sigma_- g(\rv_2,t_2,\rv_1,t_1)]
 \label{Dmag1def}
\end{align}
Denoting $(\rv,t)$ by $x$, the Green's function part
is
\begin{align}
\int_C dt_3
& d(x,x_1)
 \tr [\sigma_+g(x_1,x_2) \sigma_- g(x_2,x_3)  (\Av(t_3)\cdot\hat{\pv}) g(x_3,x_1)
 \nnr&
 +g(x_1,x_3) (\Av(t_3)\cdot\hat{\pv}) g(x_3,x_2) \sigma_- g(x_2,x_1)\sigma_+]
 d(x_2,x')
 \nnr
 &=
 d(x,x_1)
 [\tilde{\Lambda}_-(x_1,x_2)  + \tilde{\Lambda}_+(x_2,x_1)]
 d(x_2,x')
\end{align}
where (subscripts $\pm$ denotes electron spin)
\begin{align}
 \tilde{\Lambda}_-(x_1,x_2)  & \equiv g_-(x_1,x_2)(gAg)_+(x_2,x_1) \nnr
 \tilde{\Lambda}_+(x_2,x_1) & \equiv g_+(x_2,x_1) (gAg)_-(x_1,x_2)
\end{align}
and
$gAg(x_2,x_1) \equiv \int_C dt_3 g(x_2,x_3)  (\Av(t_3)\cdot\hat{\pv}) g(x_3,x_1)$
behaves as composite propagators.
The lesser component, $D^{(1)<}$, for Eq. (\ref{jmdef})
is calculated  using
\begin{align}
\int_C dt_1 &  \int_C dt_2
 [d(x,x_1)
 [\tilde{\Lambda}_-(x_1,x_2)  + \tilde{\Lambda}_+(x_2,x_1)]
 d(x_2,x')]^<\nnr
 &=
 \int_{-\infty}^\infty  dt_1 \int_{-\infty}^\infty dt_2
 \sum_{\alpha\beta} (-)^2\alpha \beta \lt[  d^{-\alpha}(x,x_1)
 [\tilde{\Lambda}_-^{\alpha\beta}(x_1,x_2)  + \tilde{\Lambda}_+^{\beta\alpha}(x_2,x_1) \rt] d^{\beta +}(x_2,x')
\end{align}
Here superscripts such as in $d^{\alpha\beta}$ ($\alpha,\beta=\pm$) denotes the time contour:
$d^{--}(t,t_1)=d(t\in C_\rightarrow, t_1\in C_\rightarrow)$,
$d^{-+}(t,t_1)=d(t\in C_\rightarrow, t_1\in C_\leftarrow)$, etc.
The result is (suppressing the time integration)
\begin{align}
[\int_C dt_1  \int_C dt_2
 d(x,x_1)
 \tilde{\Lambda}_-(x_1,x_2)
 d(x_2,x')]^<
 &=
\frac{1}{2}(d^\ret\tilde{\Lambda}_-^{\rm K}d^\adv +d^\ret \tilde{\Lambda}_-^\ret d^{\rm K}
  +d^{\rm K}\tilde{\Lambda}_-^\adv  d^\adv)\nonumber \\
[\int_C dt_1   \int_C dt_2
 d(x,x_1)
 \tilde{\Lambda}_+(x_2,x_1)
 d(x_2,x')]^<
 &=
\frac{1}{2}(d^\ret\tilde{\Lambda}_+^{\rm K}d^\adv +d^\ret \tilde{\Lambda}_+^\adv d^{\rm K}
  + d^{\rm K}\tilde{\Lambda}_+^\ret  d^\adv)
 \end{align}
Fourier representation of $\tilde{\Lambda}$ is (using $A(t) \propto e^{-i\Omega t}$)
\begin{align}
 \tilde{\Lambda}_\sigma(x_1,x_2)
 &=
 \sum_{\kv\kv'}\sum_{\Omega\omega\omega'} e^{-i(\omega-\omega')(t_1-t_2)} e^{-i(\kv-\kv')\cdot(\rv_1-\rv_2)} e^{-i\Omega t_2} (\Av(\Omega)\cdot{\kv})
\Lambda_{k,\omega+\Omega,\omega,-\sigma}^{k',\omega',\sigma}
 \end{align}
where
\begin{align}
\Lambda_{k,\omega+\Omega,\omega,-\sigma}^{k',\omega',\sigma}
\equiv
g_{\kv',\omega',\sigma} g_{\kv,\omega+\Omega,-\sigma} g_{\kv,\omega,-\sigma}
\label{Lambdadef}
 \end{align}

The magnon current therefore reads
 \begin{align}
 j_{{\rm m},i}^{(1)} &= -\frac{e}{2m}J\Deltasd^2 \sum_{p\nu} \sum_{k\omega} \sum_{\Omega}p_i k_jA_j(\Omega)
 \biggl[ \lt\{ d_{p\nu}^\ret d_{p\nu}^\adv  [\Lambda^{k+p,\omega+\nu,-}_{k,\omega+\Omega,\omega,+}]^{\rm K} +(rKr)+(Kaa) \rt\}
 \nnr &
 + \lt\{ d_{p\nu}^\ret d_{p\nu}^\adv  [\Lambda^{k-p,\omega-\nu,+}_{k,\omega+\Omega,\omega,-}]^{\rm K} +(rKa)+(Kar) \rt\} \biggr]
 \label{jm1}
\end{align}
Here the external frequency $\Omega$ is neglected in the magnon frequencies ($\nu+\Omega \sim \nu$) focusing the Fermi surface (excitation) contribution.

In Eq. (\ref{jm1}), (rKr)+(Kaa) terms are, using
$d_{p\nu}^{\rm K}=-(1+2n_\nu)(d_{p\nu}^{\adv} - d_{p\nu}^{\ret})$,
\begin{align}
 d_{p\nu}^\ret d_{p\nu}^{\rm K}[\Lambda^{k+p,\omega+\nu,-}_{k,\omega+\Omega,\omega,+}]^\ret
 & + d_{p\nu}^{\rm K} d_{p\nu}^{\adv}[\Lambda^{k+p,\omega+\nu,-}_{k,\omega+\Omega,\omega,+}]^\adv
 \nnr
&=
-(1+2n_\nu)[ d_{p\nu}^\ret (d_{p\nu}^{\adv} - d_{p\nu}^{\ret}) [\Lambda^{k+p,\omega+\nu,-}_{k,\omega+\Omega,\omega,+}]^\ret
+ (d_{p\nu}^{\adv} - d_{p\nu}^{\ret})d_{p\nu}^{\adv}[\Lambda^{k+p,\omega+\nu,-}_{k,\omega+\Omega,\omega,+}]^\adv]
\nnr
&\simeq (1+2n_\nu) d_{p\nu}^\ret d_{p\nu}^{\adv} \lt[ [\Lambda^{k+p,\omega+\nu,-}_{k,\omega+\Omega,\omega,+}]^\adv -[\Lambda^{k+p,\omega+\nu,-}_{k,\omega+\Omega,\omega,+}]^\ret \rt]
\end{align}
where contributions $(d_{p\nu}^\ret )^2$ and  $(d_{p\nu}^\adv )^2$ are neglected as they are smaller compared to  $d_{p\nu}^\ret d_{p\nu}^{\adv}$ after $\nu$ summation.
The magnon current in this approximation reads
 \begin{align}
 j_{{\rm m},i}^{(1)}
 &= -\frac{e}{2m}J\Deltasd^2 \sum_{p\nu} \sum_{k\omega} \sum_{\Omega} p_i k_jA_j(\Omega) d_{p\nu}^\ret d_{p\nu}^\adv
 \lt[ [\Lambda^{k+p,\omega+\nu,-}_{k+,\omega+\Omega,\omega}]^{\rm K}+(1+2n_\nu)\lt[ [\Lambda^{k+p,\omega+\nu,-}_{k+,\omega+\Omega,\omega}]^{\adv}- [\Lambda^{k+p,\omega+\nu,-}_{k+,\omega+\Omega,\omega}]^{\ret} \rt]
 \rt. \nnr & \lt. +   [\Lambda^{k-p,\omega-\nu,+}_{k-,\omega+\Omega,\omega}]^{\rm K} -(1+2n_\nu) \lt[
 [\Lambda^{k-p,\omega-\nu,+}_{k-,\omega+\Omega,\omega}]^{\adv} - [\Lambda^{k-p,\omega-\nu,+}_{k-,\omega+\Omega,\omega}]^{\ret} \rt]
 \rt]
 \label{jm2}
\end{align}
The Keldysh, advanced and retarded components of $\Lambda$ are (in the suppressed notation)
 \begin{align}
\Lambda^{\rm K} &=
g^{<}(gg)^{>}+g^>(gg)^<
 \nnr
\Lambda^\adv &= g^{\adv}(gg)^{<}+g^{<}(gg)^\ret =  g^{\adv}(gg)^{>}+g^{>}(gg)^\ret \nnr
\Lambda^\ret &= g^{\ret}(gg)^{<}+g^{<}(gg)^\adv =  g^{\ret}(gg)^{>}+g^{>}(gg)^\adv
\label{LambdaKar}
\end{align}
Focusing on the Fermi surface (excitation) contribution, we neglect terms containing only the retarded or the advanced Green's functions, to obtain
 \begin{align}
\Lambda^{\rm K} & \simeq
 (f_\omega-f_{\omega'})(2f_{\omega''}-1) (g^\adv-g^\ret)g^\ret g^\adv \nnr
&
 - [f_{\omega''}(f_{\omega'}-1)+f_{\omega'}(f_{\omega''}-1)]  g^\ret(g g)^\adv
 - [f_{\omega''}(f_{\omega}-1)+f_{\omega}(f_{\omega''}-1)] g^\adv (g g)^\ret
 \nnr
\Lambda^{\adv}
&\simeq
f_{\omega''}g^\adv(gg)^\ret + g^\adv\lt[(f_\omega-f_{\omega'})g^\ret g^\adv -f_\omega(gg)^\ret \rt]
\nnr
\Lambda^{\ret}
&\simeq - f_{\omega''}g^\ret(gg)^\adv + g^\ret\lt[(f_\omega-f_{\omega'})g^\ret g^\adv +f_{\omega'}(gg)^\adv \rt]
 \end{align}
 where $\Lambda$ of Eq. (\ref{Lambdadef}) is simply denoted by $g_{\omega''}g_{\omega'}g_{\omega}$ (with $\omega''=\omega\pm\nu$, $\omega'-\omega+\Omega$).
Using the Fermi surface approximation  for the two Greens function,   $(gg)^<\simeq (f_\omega-f_{\omega'})g^\ret g^\adv$, we obtain
 \begin{align}
\Lambda^{\rm K} & \simeq
 (2f_{\pm\nu}-1) (\Lambda^{\adv}-\Lambda^{\ret}) \nnr
\Lambda^{\adv}
&\simeq
-\Omega f'(\omega)g^\adv g^\ret g^\adv
\nnr
\Lambda^{\ret}
&\simeq -\Omega f'(\omega)g^\ret g^\ret g^\adv
\label{LamKFS}
\end{align}
Considering low temperature, $f'(\omega)\simeq -\delta(\omega)$, we obtain (using $2f_{-\nu}-1=1-2f_\nu$)
 \begin{align}
 j_{{\rm m},i}^{(1)}
 &\simeq
 - \frac{e}{2\pi m}J\Deltasd^2 \sum_{p\nu} \sum_{k\omega} \sum_{\Omega} p_i k_j \Omega A_j(\Omega)| d_{p\nu}^\adv|^2
 \sum_{\pm}(\pm)
(f_{\nu}+n_\nu) \lt[ [\Lambda^{k\pm p,\pm\nu,\mp}_{k,\pm,0,0}]^{\adv}-[\Lambda^{k\pm p,\pm\nu,\mp}_{k,\pm,0,0}]^{\ret}
 \rt]
 \label{jm3}
\end{align}
Using $ f_{\nu}+n_\nu=2n_\nu(1-f_\nu)$, and $i\Omega A=E$, we obtain the final result of
 \begin{align}
 j_{{\rm m},i}^{(1)}
 &\simeq
 -\frac{2e}{\pi m} {J\Deltasd^2}{\tau_{\rm m}} \sum_{p\nu} \sum_{k\omega} \sum_{\Omega} p_i k_jE_j \Im[ d_{p\nu}^\adv]
 \sum_{\pm}(\pm)
n_\nu (1-f_{\nu}) \Im[g_{k\pm p,\pm\nu,\mp}^\adv]|g_{k,\pm}^{\adv}|^2
\nnr &\equiv \sigma_{{\rm m},i,j}^{(1)} E_{j}
 \label{jm4}
\end{align}
We therefore confirm the correct relation between the force and velocity of magnon,
$ \sigma_{{\rm m},ij}^{(1)}= {\cal F}_{ij}^{(1)} \frac{\tau_{\rm m}}{m_{\rm m}}$, where $m_{\rm m}=(2JS)^{-1}$ is the effective magnon mass.

\subsection{Magnon scattering contribution}

\begin{figure}
 \includegraphics[width=0.5\hsize]{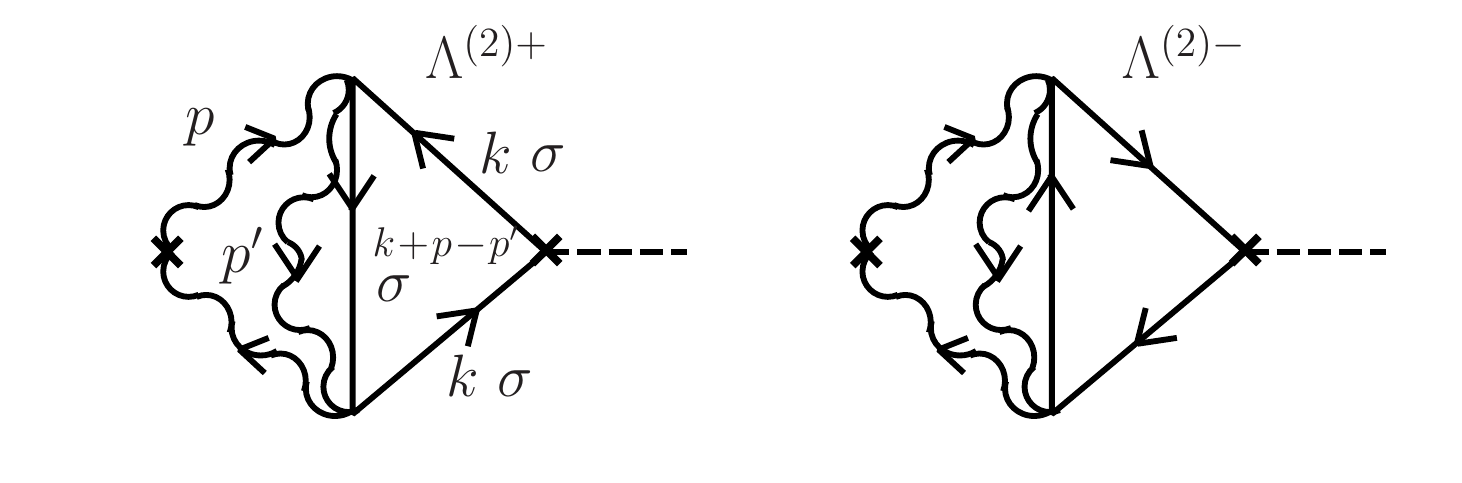}
 \caption{
Feynman diagrams representing the magnon current (left vertex) induced by an electric field (denoted by a dotted line) due to the magnon scattering by the $sd$ exchange interaction. Wavy and solid lines denote magnon and electron propagators, respectively. $\sigma$ denotes electron spin.
\label{FIGmagj2}}
\end{figure}
Here the second-order magnon coupling to electrons representing the magnon scattering arising from the $sd$ exchange interaction is studied.
This contribution was not studied in Ref. \cite{Miura12}.
The contribution to the magnon Green's function defined on the time contour, $D^{(2)}(\rv,t,\rv',t')$, is (Fig. \ref{FIGmagj2})
\begin{align}
 D^{(2)} & (\rv,t,\rv',t')
  =
  i\frac{e}{m}\frac{\Deltasd^2}{S^2} \int_C dt_1 \int_C dt_2 \int_C dt_3
  \sum_{r_1r_2r_3} d(\rv_2,t_2,\rv',t') d(\rv,t,\rv_1,t_1)  A_j(t_3) \nnr
 & \times
 \tr [\sigma_z \Pi^{(+)}({\rv_1,t_1},\rv_2,t_2) \sigma_z g(\rv_2,t_2,\rv_3,t_3) \hat{p}_jg(\rv_3,t_3,\rv_1,t_1) \nnr&
 +\sigma_z g({\rv_1,t_1},\rv_3,t_3) \hat{p}_j g(\rv_3,t_3,\rv_2,t_2) \sigma_z \Pi^{(-)}(\rv_2,t_2,\rv_1,t_1)]
\label{Dmag2def}
 \end{align}
 where $\Pi^{(\pm)}$ are magnon-electron pair propagators defined in Eq. (\ref{Pidefs}) (suffixes are suppressed in Eq. (\ref{Dmag2def})).
As the expression of Eq. (\ref{Dmag2def}) has the same structure as Eq. (\ref{Dmag1def}), we immediately obtain
the magnon current as (see  Eq. (\ref{jm1}))
 \begin{align}
 j_{{\rm m},i}^{(2)} &= -\frac{e}{2mS}J\Deltasd^2 \sum_{p\nu}\sum_{k\omega\Omega} p_i k_jA_j(\Omega) \nonumber\\
 & \times
 \lt[ \lt\{ d_{p\nu}^\ret d_{p\nu}^\adv  [\Lambda^{(2+),k+p,\omega+\nu}_{k,\omega+\Omega,\omega}]^{\rm K} +(rKr)+(Kaa) \rt\}
 + \lt\{ d_{p\nu}^\ret d_{p\nu}^\adv  [\Lambda^{(2-),k-p,\omega-\nu}_{k,\omega+\Omega,\omega}]^{\rm K} +(rKa)+(Kar) \rt\}\rt]
 \label{jm21}
\end{align}
where (($\pv'$ is the wave vector of the magnon in the pair propagators)
 \begin{align}
 \Lambda^{(2+),k+p,\omega+\nu}_{k,\omega+\Omega,\omega}
 &= \sum_{\sigma} \Pi_{\kv+\pv,\omega+\nu}^{(+),(\kv+\pv-\pv',\sigma,\pv')} g_{\kv,\omega+\Omega,\sigma} g_{\kv,\omega,\sigma}
 \nnr
 \Lambda^{(2-),k-p,\omega-\nu}_{k,\omega+\Omega,\omega}
 &= \sum_{\sigma}\Pi_{\kv-\pv,\omega-\nu}^{(-),(\kv-\pv+\pv',\sigma,\pv')} g_{\kv,\omega+\Omega,\sigma} g_{\kv,\omega,\sigma}
\end{align}
Using the Fermi surface approximation as in Eq. (\ref{LamKFS}),
\begin{align}
\Lambda^{(2){\rm K}}
& \simeq
 (2f_{\pm\nu}-1) (\Lambda^{(2)\adv}-\Lambda^{(2)\ret}) \nnr
\Lambda^{(2)\adv}
&\simeq (f_\omega-f_{\omega'})
\Pi^\adv g^\ret g^\adv \simeq -\Omega f'(\omega)\Pi^\adv g^\ret g^\adv
\nnr
\Lambda^{(2)\ret}
&\simeq -\Omega f'(\omega)\Pi^\ret g^\ret g^\adv
\label{LamK2FS}
\end{align}
we  obtain
 \begin{align}
 j_{{\rm m},i}^{(2)}
 &\simeq  -\frac{2e}{\pi mS}J\Deltasd^2 \sum_{pp'k} \sum_{\nu}p_i k_jE_j | d_{p\nu}^\adv|^2
 \sum_{\pm}(\pm) \sum_{\sigma}
n_\nu (1-f_{\nu}) \Im[{\Pi_{\kv\pm\pv,\pm\nu}^{(\pm),(\kv\pm\pv-\pv',\sigma,\pv')}}^\adv]|g_{k,\sigma}^{\adv}|^2
\label{jm2result}
\end{align}
Here the magnon-electron pair propagator has an external frequency of $\pm\nu$ (external magnon frequency), and its imaginary part reads  (see Eq. (\ref{ImPi2result}) for the case of $\nu=0$)
\begin{align}
 \Im \Pi_{k\pm p,\sigma, \pm\nu}^{(\pm),(k\pm(p-p'),\pm(\nu-\nu');p',\nu'),\adv}
 &=
 2 \sum_{\nu'} (n_{\nu'}+f_{\nu'-\nu})  \Im [g_{k\pm(p-p'),\pm(\nu-\nu'),\sigma}^\adv] \Im[d^\adv_{p',\nu'}]
\label{ImPi12}
\end{align}
 Using $(n_{\nu'}+f_{\nu'-\nu})=(1+n_{\nu'})f_{\nu'-\nu}(1-f_\nu)^{-1}$ and
 $\Im[d_{p'\nu'}^\adv]=2\eta_{\rm m}| d_{p'\nu'}^\adv|^2=(\tau_{\rm m})^{-1}|d_{p'\nu'}^\adv|^2$ ($\tau_{\rm m}$ is the relaxation time for magnon),
we obtain
 \begin{align}
 j_{{\rm m},i}^{(2)}
 &=  -\frac{2e}{\pi mS}\frac{J\Deltasd^2} {\tau_{\rm m}}\sum_{pp'k} p_i k_jE_j  \Im[d_{p\nu}^\adv] \Im[ d_{p'\nu'}^\adv]
 \sum_{\pm}(\pm)\sum_{\nu\nu'}\sum_{\sigma}
n_\nu (1+n_{\nu'})f_{\nu'-\nu} \Im[g_{k\pm (p-p'),\pm(\nu-\nu'),\sigma}^\adv]|g_{k,\sigma}^{\adv}|^2
 \label{jm24}
 \end{align}
 confirming the relation between the corresponding force density and current density,
$ j_{{\rm m},i}^{(2)}={\cal F}^{(2)}\frac{\tau_{\rm m}}{m_{\rm m}}$.

\input{magnonforce.bbl}
\end{document}

%% file: magnonforce.bbl
%